\documentclass[aps,prd,twocolumn,superscriptaddress,floatfix]{revtex4-2}

\usepackage{amsmath,amssymb}
\usepackage{graphicx}
\usepackage{xcolor}
\usepackage{hyperref}
\usepackage{physics}
\usepackage{bm}
\usepackage{multirow}
\usepackage{booktabs}
\usepackage{comment}
\usepackage{soul}  % for \st{} strikethrough in red corrections

\hypersetup{
	colorlinks=true,
	linkcolor=blue,
	filecolor=magenta,      
	urlcolor=cyan,
	citecolor=blue,
}

\begin{document}
	
	\title{Quantum Backreaction in Effective Brans-Dicke Bianchi I Cosmology}
	
	\author{Hector Hernandez-Hernandez}
	\email{hhernandez@uach.mx}
	\affiliation{Universidad Autonoma de Chihuahua, Facultad de Ingenieria,
		Nuevo Campus Universitario, Chihuahua 31125, Mexico}
	\affiliation{
		Universidad Autonoma Metropolitana- Cuajimalpa, Departamento de Matematicas Aplicadas y Sistemas, Vasco de Quiroga 4871, 05348, Mexico}
	
	\author{Gustavo Sanchez-Herrera}
	\email{cbi2233805002@xanum.uam.mx}
	\affiliation{Universidad Autonoma Metropolitana- Iztapalapa, Departamento de Fisica, 
		Ciudad de Mexico, 09340, Mexico}
	
	%\date{\today}
	
	\begin{abstract}
		We investigate the effective quantum evolution of the Bianchi type~I 
		cosmological model within the Brans-Dicke framework, using an 
		effective Hamiltonian approach including expectation values, quantum 
		dispersions, and cross-correlation terms between different degrees of 
		freedom.  We show that cross-correlation terms are essential for a physically consistent 
		effective dynamics: neglecting them leads to spurious divergences, 
		violation of the Heisenberg uncertainty relations, and 
		unphysical behavior. For $\omega < -3/2$, where bouncing solutions exist 
		already classically, quantum backreaction smooths the bounce, suppresses shear anisotropy, and the energy density peak is slightly enhanced and remains well-behaved, in contrast to the unphysical divergences that arise when cross-correlations are neglected. For the conformally invariant case 
		$\omega = -3/2$, quantum corrections accelerate the approach to de~Sitter 
		expansion while enhancing shear 
		anisotropy. When correlations are included, small-amplitude damped 
		oscillations appear shortly after the bounce, which we interpret as 
		quantum remnant effects encoding Planck-scale correlation 
		information between gravitational and scalar field degrees of freedom. 
		The effective energy density remains finite for both regimes. Our 
		results demonstrate that cross-correlations carry crucial quantum 
		information influencing cosmological dynamics, with implications for quantum 
		gravity phenomenology, inflationary model building, and 
		primordial observational signatures.
	\end{abstract}
	
	
	\maketitle
	
	%%%%%%%%%%%%%%%%%%%%%%%%%%%%%%%
	\section{Introduction}
	\label{sec:intro}
	%%%%%%%%%%%%%%%%%%%%%%%%%%%%%%%
	
	General Relativity (GR)~\cite{einstein1916grundlagen}, has achieved remarkable empirical success 
	over the past century. Its predictions, from the anomalous precession 
	of Mercury's perihelion and gravitational light 
	deflection to the existence of black 
	holes and gravitational 
	waves~\cite{abbott2016observation}, have been repeatedly 
	confirmed with increasing precision. Nevertheless, several fundamental 
	challenges indicate that GR cannot be the complete description of 
	gravitational phenomena.
	Most prominently, GR predicts its own breakdown at cosmological 
	singularities, both the Big Bang and black hole interiors, where 
	spacetime curvature diverges~\cite{hawking1970singularities}. 
	These singularities signal the need for a quantum theory of gravity. 
	Additionally, observations of galactic rotation 
	curves~\cite{rubin1980rotational}, large-scale structure 
	formation~\cite{abbott2018dark}, and the accelerated cosmic 
	expansion~\cite{riess1998observational} 
	require either the introduction of dark matter and dark energy, or 
	modifications to gravitational dynamics itself. These puzzles have 
	motivated extensive investigations of both quantum gravity and modified 
	gravity theories.
	
	Among the earliest and most influential generalizations of GR is the 
	Brans-Dicke (BD) theory~\cite{brans1961mach}, developed in 1961 to 
	incorporate Mach's principle into gravitational physics. In BD theory, 
	Newton's gravitational constant $G$ is promoted to a dynamical scalar 
	field $\phi$, with $G \propto \phi^{-1}$. The theory contains a 
	dimensionless parameter $\omega$ that governs the strength of the 
	coupling between the scalar field and gravity. Solar system tests 
	constrain $\omega > 40{,}000$ for standard BD 
	theory~\cite{will2014confrontation,avilez2014cosmological}, although these constraints 
	can be evaded in modified versions~\cite{faraoni2004scalar}.
	
	Despite these tight observational bounds, the regime 
	$\omega < -3/2$ presents compelling theoretical interest. First, $\omega=-3/2$ BD theory 
	is dynamically equivalent to Palatini $f(R)$ 
	gravity~\cite{sotiriou2010f}, an observationally 
	motivated class of modified gravity theories constrained by 
	cosmological data. Second, the low-energy effective action of string 
	theory takes the form of a BD action in the string 
	frame~\cite{gasperini2003pre,veneziano1991scale}, making the 
	$\omega \leq -3/2$ sector directly relevant to pre-big-bang 
	cosmology. Third, recent proposals to resolve 
	the Hubble and $S_8$ 
	tensions~\cite{di2021realm} invoke BD-like 
	mechanisms~\cite{bisabr2024brans} that require 
	understanding the quantum structure of BD bouncing solutions. Specifically, BD modifications to early dark energy require $\omega < 0$, and the quantum bounce structure we study here is directly relevant to the viability of such proposals. Most 
	importantly, $\omega < -3/2$ BD theory avoids singularities already at 
	the classical level through energy condition 
	violation~\cite{almeida2021quantum,batista2001remark}, providing a unique 
	theoretical laboratory for disentangling classical modified-gravity 
	effects from genuine quantum corrections to the bounce. The results 
	obtained here, particularly regarding the essential role of 
	cross-correlations, are expected to hold for any anisotropic quantum 
	cosmological model with coupled degrees of freedom, regardless of the 
	specific value of $\omega$.
	
	Moreover, violation of energy conditions~\cite{almeida2021quantum} 
	potentially allow for nonsingular bouncing cosmologies without 
	exotic matter~\cite{batista2001remark} and accelerated expansion mechanisms~\cite{hrycyna2014dynamics}. 
	Whether these applications require $\omega < 0$ remains an active area 
	of investigation.
	
	A fundamental hypothesis for quantum gravity is the resolution of 
	classical singularities through quantum effects. Loop Quantum Cosmology 
	(LQC), the cosmological sector of Loop Quantum Gravity 
	(LQG)~\cite{ashtekar2011loop}, has demonstrated 
	singularity replacement by quantum bounces in the homogeneous, 
	isotropic (FLRW) model~\cite{bojowald2001absence,ashtekar2006quantum}. The 
	mechanism relies on quantum geometry effects becoming dominant near the 
	Planck scale, generating an effective repulsive force that prevents 
	collapse to infinite density.
	
	Extensions to anisotropic models, particularly Bianchi cosmologies, 
	are crucial for several reasons. Some of the most important are 1) the 
	Belinskii-Khalatnikov-Lifshitz (BKL) 
	conjecture~\cite{lifshitz1963investigations,belinskii1970oscillatory} proposes that 
	generic spacetimes near singularities exhibit chaotic Bianchi-type 
	dynamics, making anisotropic models phenomenologically essential, 2) 
	anisotropies may leave observational imprints in the cosmic microwave 
	background~\cite{agullo2013extension}, 3) technical complexities of 
	anisotropic models test the robustness of quantum gravity approaches. 
	Loop quantum Bianchi~I models have been studied in 
	GR~\cite{ashtekar2009loop}, demonstrating 
	singularity resolution with bounded curvature invariants. However, 
	full quantum evolution equations remain challenging to solve 
	analytically for systems without reduced symmetry.
	
	For complex quantum systems where the full Schr\"{o}dinger or 
	Wheeler-DeWitt equation is intractable, effective quantum theories 
	provide a powerful alternative~\cite{bojowald2006effective,bojowald2012quantum}. 
	Rather than solving for the complete wavefunction $\Psi$, one works 
	directly with expectation values $\langle \hat{O} \rangle$, quantum 
	dispersions $\Delta(\hat{O}^2) = \langle (\hat{O} - \langle \hat{O} 
	\rangle)^2 \rangle$, and correlations between observables. The 
	effective Hamiltonian approach, developed systematically 
	in~\cite{bojowald2006effective,bojowald2007large,bojowald2012quantum}, expands 
	the expectation value of the quantum Hamiltonian in powers of quantum 
	moments:
	\begin{equation}
		H_{\text{eff}} = \langle \hat{H} \rangle = H_{\text{cl}}[\langle 
		\hat{q} \rangle, \langle \hat{p} \rangle] + \sum_{n=2}^{\infty} 
		H^{(n)}[\Delta(q^a p^b)],
	\end{equation}
	where $H^{(n)}$ contains contributions from $n$-th order moments. 
	Truncating at a given order yields a finite-dimensional extended phase 
	space including expectation values up to that order in moments.
	
	This approach has successfully reproduced known results in isotropic 
	quantum cosmology~\cite{bojowald2007effective} while enabling studies of 
	more complex scenarios: anisotropic Bianchi 
	models~\cite{hernandez2024singularity}, inhomogeneous 
	cosmologies~\cite{gomar2015gauge} and general quantum 
	systems~\cite{chacon2024effective,aragon2020effective}.
	
	A critical aspect often overlooked in effective approaches is the role 
	of cross-correlation terms, quantum moments coupling different degrees 
	of freedom, such as $\Delta(q_i p_j)$ with $i \neq j$ or 
	$\Delta(q_i \phi)$ between gravitational and matter variables. Recent 
	work has demonstrated that correlations can substantially modify the 
	dynamics~\cite{bojowald2021canonical,hernandez2023semiclassical}: 
	neglecting correlations may miss crucial quantum effects because they 
	encode nonlocal quantum information and, in some systems, correlations 
	prevent unphysical behavior. We demonstrate here, both 
	analytically and numerically, that cross-correlations are dynamically 
	generated by the Hamiltonian interaction structure and that neglecting them violates the Heisenberg uncertainty 
	relations and produces spurious divergences, establishing them as an 
	essential feature of anisotropic quantum cosmology rather than a 
	perturbative correction.
	
	In this paper, we investigate the effective quantum dynamics of the 
	Bianchi~I cosmological model within Brans-Dicke theory, with 
	particular focus on studying how quantum corrections modify classical 
	bouncing solutions (quantum backreaction); analysing the quantitative 
	impact of cross-correlations on dynamics and whether they are essential 
	for consistency; and investigating the coupling parameter dependence and 
	the physical interpretation of our findings such as the nature of 
	post-bounce oscillations and how quantum remnants manifest.
	
	This article is structured as follows. In 
	section~\ref{sec:classical} we review the Hamiltonian formulation of 
	Brans-Dicke theory for Bianchi~I spacetime and identify the physical observables such as the Hubble parameter, the energy density, and shear. We also present classical 
	dynamics for both coupling constant regimes (all technical details in 
	Appendix~\ref{app:adm}). In section~\ref{sec:effective} we develop 
	the effective quantum formalism and derive the second-order effective 
	Hamiltonian, with a discussion of the truncation validity. In 
	section~\ref{sec:numerical} we present numerical results, emphasizing 
	the crucial role of cross-correlations. In 
	section~\ref{sec:discussion} we compare our findings with previous 
	work, discuss physical interpretation and summarize results and outline future directions. Appendix~\ref{app:adm} contains ADM formulation details. Appendix~\ref{app:eom} gives complete equations of motion.
	
	\section{Classical Brans-Dicke Bianchi I Cosmology}
	\label{sec:classical}
	
	In this section, we establish the classical framework for our quantum effective analysis. We begin with the Hamiltonian formulation of Brans-Dicke theory specialized to Bianchi type I spacetime (Sec.~\ref{subsec:hamiltonian}), then analyze the classical dynamics for two physically distinct cases: generic negative coupling $\omega < -3/2$ (Sec.~\ref{subsec:classical_generic}) and the conformally invariant case $\omega = -3/2$ (Sec.~\ref{subsec:classical_conformal}).
	
	\subsection{Hamiltonian Formulation in Ashtekar Variables}
	\label{subsec:hamiltonian}
	
	\subsubsection{Bianchi I Geometry and Canonical Variables}
	
	The Bianchi type I model describes spatially homogeneous, anisotropic, and spatially flat cosmologies. The metric takes the diagonal form
	\begin{equation}
		ds^2 = -N^2 dt^2 + a_1^2(t) dx_1^2 + a_2^2(t) dx_2^2 + a_3^2(t) dx_3^2,
		\label{eq:bianchi1_metric}
	\end{equation}
	where $N(t)$ is the lapse function and $a_i(t)$ are directional scale factors. Spatial homogeneity allows reduction to a finite-dimensional phase space after introducing a fiducial cell $\mathcal{V}_0$ with volume $V_0 = L_1 L_2 L_3$~\cite{ashtekar2009loop}.
	
	We employ Ashtekar-Barbero variables~\cite{barbero1994real}, which recast gravitational phase space in terms of an $\mathfrak{su}(2)$-valued connection $A_a^i$ and its conjugate densitized triad $E_i^a$. For Bianchi I, symmetry reduction yields~\cite{ashtekar2009loop}
	\begin{equation}
		A_a^i = c_i (L_i)^{-1} \mathring{e}_a^i, \quad
		E_i^a = p_i L_i V_0^{-1} \sqrt{\mathring{q}} \, \mathring{e}^a_i,
		\label{eq:ashtekar_vars}
	\end{equation}
	where $\mathring{e}_a^i$ is a fiducial triad with determinant $\mathring{q}$, and $(c_i, p_i)$ are time-dependent canonical variables satisfying
	\begin{equation}
		\{c_i, p_j\} = 8\pi G \gamma \delta_{ij},
		\label{eq:poisson_cp}
	\end{equation}
	with $\gamma$ the Barbero-Immirzi parameter~\cite{immirzi1997real}.
	
	The directional scale factors are related to $p_i$ by (choosing positive orientation and $L_i = 1$)
	\begin{equation}
		a_1 = \sqrt{\frac{p_2 p_3}{p_1}}, \quad
		a_2 = \sqrt{\frac{p_1 p_3}{p_2}}, \quad
		a_3 = \sqrt{\frac{p_1 p_2}{p_3}}.
		\label{eq:scale_factors}
	\end{equation}

	The mean scale factor is $a = (a_1 a_2 a_3)^{1/3} = (p_1 p_2 p_3)^{1/6} \equiv p^{1/6}$.
	
	Physical observables include directional Hubble parameters
	\begin{equation}
		H_i \equiv \frac{\dot{a}_i}{a_i} = \frac{1}{2}\left( \sum_{j \neq i} \frac{\dot{p}_j}{p_j} - \frac{\dot{p}_i}{p_i} \right),
		\label{eq:hubble_directional}
	\end{equation}
	and the shear scalar, measuring anisotropy:
	\begin{equation}
		\Sigma^2 = \frac{1}{3}\sum_{i<j} (H_i - H_j)^2.
		\label{eq:shear}
	\end{equation}
	
	\subsubsection{Brans-Dicke Action and Hamiltonian Constraint}
	
	The Brans-Dicke action is~\cite{brans1961mach,faraoni2004scalar}
	\begin{equation}
		S_{\text{BD}} = \frac{1}{16\pi} \int d^4x \sqrt{-g} \left[ \phi R - \frac{\omega}{\phi} g^{\mu\nu} \partial_\mu \phi \partial_\nu \phi \right],
		\label{eq:bd_action}
	\end{equation}
	where $\phi$ is the Brans-Dicke scalar (related to the gravitational constant by $G_{\text{eff}} = 1/\phi$) and $\omega$ is the dimensionless coupling constant. We work in units where $16\pi G = 1$ and set $\gamma = 1$ for simplicity. The Barbero--Immirzi parameter $\gamma$ can take any positive real value; here $\gamma = 1$ is chosen purely for notational simplicity and does not affect the qualitative dynamics. All results can be restored to general $\gamma$ by the replacements $c_i \to c_i/\gamma$ and $p_i \to \gamma p_i$ in the equations below.
	
	Following the procedure detailed in Appendix~\ref{app:adm}, the Hamiltonian constraint for Bianchi I in Ashtekar variables becomes~\cite{sharma2025quantum}:
	
	\paragraph{Case I: $\omega \neq -3/2$ (Generic Coupling).}
	
	For $\omega < -3/2$, the Hamiltonian constraint is
	\begin{equation}
		\mathcal{H}_{\text{BD}} = -\frac{N}{\gamma^2 \phi \sqrt{p}} \left( S - \zeta T^2 \right) \approx 0,
		\label{eq:hamiltonian_generic}
	\end{equation}
	where we defined
	\begin{subequations}
		\begin{align}
			S &\equiv c_1 c_2 p_1 p_2 + c_1 c_3 p_1 p_3 + c_2 c_3 p_2 p_3, \label{eq:S_def} \\
			T &\equiv \sum_{i=1}^3 c_i p_i + \gamma \phi p_\phi, \label{eq:T_def} \\
			\zeta &\equiv \frac{1}{3 + 2\omega}, \label{eq:zeta_def}
		\end{align}
	\end{subequations}
	and $(\phi, p_\phi)$ are canonical variables for the scalar field with $\{\phi, p_\phi\} = 1$.
	
	\paragraph{Case II: $\omega = -3/2$ (Conformal Invariance).}
	
	For $\omega = -3/2$, the term proportional to $T^2$ diverges and the theory becomes conformally invariant~\cite{lin2023conformally}, requiring a modified constraint structure. The Hamiltonian constraint becomes in this case
	\begin{equation}
		\mathcal{H}_{\text{BD}}^c = -\frac{N}{\gamma^2 \phi \sqrt{p}} S + \frac{\lambda}{\gamma} \left( \sum_{i=1}^3 c_i p_i - \phi p_\phi \right) \approx 0,
		\label{eq:hamiltonian_conformal}
	\end{equation}
	where $\lambda$ is a Lagrange multiplier that contains no new physical information and imposes an additional constraint
	\begin{equation}
		\mathcal{C}_S = \sum_{i=1}^3 c_i p_i - \phi p_\phi \approx 0.
		\label{eq:conformal_constraint}
	\end{equation}
	The constraint (\ref{eq:conformal_constraint}), associated with the conformal invariance characteristic of the special case $\omega = -3/2$, constitutes a first class constraint \cite{olmo2011hamiltonian}. Since the total Hamiltonian is constructed as a linear combination of all first class constraints (each generating an independent gauge symmetry), this structure implies the appearance of two Lagrange multipliers $N$ and $\lambda$ in the theory. Note that the second term in Eq.~\eqref{eq:hamiltonian_conformal} lacks the lapse factor $N$: this is because $\mathcal{C}_S$ arises as an independent primary first-class constraint via the Dirac--Bergmann algorithm, with its own independent Lagrange multiplier $\lambda$ distinct from $N$. The total Hamiltonian takes the form $\mathcal{H}_{\text{BD}}^c = N\,\mathcal{H}_{\text{BD}}^{(0)} + \lambda\,\mathcal{C}_S$, where the two multipliers generate separate gauge transformations, and $\mathcal{H}_{\text{BD}}^{(0)}=-\frac{1}{\gamma^2 \phi \sqrt{p}} S$
	
	%%%%%%%%%%%%%%%%%%%%%%%%%%%%%%%%%%%%%%%%%%%%%%%%%%%%%%%%%%
	\subsubsection{Physical Energy Densities}
	
	To identify the physical energy density, we compare the BD Hamiltonian constraints with the standard GR Friedmann structure. In the isotropic limit where all $c_i = \mathcal{C}$ and $p_i = \mathcal{P}$, the GR Hamiltonian constraint is~\cite{bojowald2011quantum}
	\begin{eqnarray}\label{TotalGravitational&MatterConstraint}
		\mathcal{H}= -\frac{3}{8 \pi G \gamma^{2}} \mathcal{C}^{2}\sqrt{|\mathcal{P}|} + \mathcal{H}_{\text{matter}} \approx 0,
	\end{eqnarray}
	where the vanishing of the constraint gives the Friedmann equation $H^2 = (8\pi G/3)\rho$, with the matter Hamiltonian reading $\mathcal{H}_{\text{matter}} = \mathcal{P}^{3/2}\rho$. For the Bianchi I case, the matter sector generalizes to $\mathcal{H}_{\text{matter}} = \sqrt{p_1 p_2 p_3}\,\rho$. By identifying the second (positive) term in constraints~\eqref{eq:hamiltonian_generic} and~\eqref{eq:hamiltonian_conformal} with this matter contribution, we can read off the effective energy density responsible for the cosmic bounce:
	\begin{equation}
		\rho' = \begin{cases}
			\dfrac{\zeta T^{2}}{\gamma^2 \phi p} , & \omega \neq -3/2 \\[6pt]
			\dfrac{\lambda}{\gamma \sqrt{p}} \left( \sum_{i=1}^3 c_i p_i - \phi p_\phi \right), & \omega = -3/2
		\end{cases}
		\label{eq:energy_density}
	\end{equation}
	
	We can check this identification by taking the isotropic limit $c_i \to \mathcal{C}$, $p_i \to \mathcal{P}$ in the constraints above with the canonical energy density $\rho = p_\phi^2/(2a^6)$, where $a = \mathcal{P}^{1/6}$. Inserting into constraint~\eqref{eq:hamiltonian_generic} and \eqref{eq:hamiltonian_conformal}, and requiring $\mathcal{H}_{\text{BD}} \approx 0$, one obtains the modified Friedmann equation
	\begin{equation}
		H^{2} \propto \begin{cases}
			\rho \left(1+ \dfrac{6 H \gamma \phi}{\sqrt{6\gamma^{2}\phi^{2} \rho}}\right) , & \omega \neq -3/2 \\[6pt]
			\rho \left(\dfrac{3 H }{2 \phi \rho}- \dfrac{1}{\sqrt{2 \rho}}\right), & \omega = -3/2
		\end{cases}
		\label{eq:energy_density_Isotropy}
	\end{equation}
	The notation $\rho'$ in Eq.~\eqref{eq:energy_density} is used to distinguish this effective BD energy density from the canonical scalar-field density $\rho$; both are finite at the bounce and serve as the key indicators of singularity avoidance.
	
	\subsection{Classical Dynamics: Case $\omega < -3/2$}
	\label{subsec:classical_generic}
	
	\subsubsection{Equations of Motion}
	
	Hamilton's equations $\dot{f} = \{f, \mathcal{H}_{\text{BD}}\}$ with the constraint~\eqref{eq:hamiltonian_generic} yield (setting $N = \gamma = 1$):
	\begin{eqnarray}
		\label{eq:eom_generic}
		\dot{c}_1 &= \frac{1}{\phi\sqrt{p}} \left[ -c_1(c_2 p_2 + c_3 p_3 - 2\zeta T) + \frac{S - \zeta T^2}{2p_1} \right], \nonumber \\
		\dot{c}_2 &= \frac{1}{\phi\sqrt{p}} \left[ -c_2(c_1 p_1 + c_3 p_3 - 2\zeta T) + \frac{S - \zeta T^2}{2p_2} \right], \nonumber \\
		\dot{c}_3 &= \frac{1}{\phi\sqrt{p}} \left[ -c_3(c_1 p_1 + c_2 p_2 - 2\zeta T) + \frac{S - \zeta T^2}{2p_3} \right], \nonumber \\
		\dot{p}_1 &= \frac{\sqrt{p}}{\phi} \left( c_2 p_3 + c_3 p_2 - \frac{2p_1 \zeta T}{p} \right), \nonumber \\
		\dot{p}_2 &= \frac{\sqrt{p}}{\phi} \left( c_1 p_3 + c_3 p_1 - \frac{2p_2 \zeta T}{p} \right), \nonumber \\
		\dot{p}_3 &= \frac{\sqrt{p}}{\phi} \left( c_1 p_2 + c_2 p_1 - \frac{2p_3 \zeta T}{p} \right), \nonumber \\
		\dot{\phi} &= \frac{2\zeta}{\sqrt{p}} T, \nonumber \\
		\dot{p}_\phi &= -\frac{1}{\phi^2 \sqrt{p}} (S + 2\gamma \zeta \phi p_\phi T - \zeta T^2).
	\end{eqnarray}
	
	We analyze this system numerically.
	
	\subsubsection{Bouncing Solutions and Initial Conditions}
	
	For $\omega < -3/2$, the parameter $\zeta < 0$, and the term $-\zeta T^2 > 0$ in the Hamiltonian constraint~\eqref{eq:hamiltonian_generic} can dominate near small volumes, generating an effective repulsive force. This leads to bouncing solutions where scale factors reach a minimum (the bounce) and then re-expand~\cite{almeida2021quantum,batista2001remark}.
	
	To explore this regime, we choose $\omega = -5$ (giving $\zeta = -1/7$) and initial conditions
	\begin{equation}
		\begin{aligned}
			c_1(0) &= 1.2, \quad p_1(0) = 1.1, \\
			c_2(0) &= 1.3, \quad p_2(0) = 1.2, \\
			c_3(0) &= 1.4, \quad p_3(0) = 1.5, \\
			\phi(0) &= 10, \quad p_\phi(0) = 1.
		\end{aligned}
		\label{eq:initial_conditions_generic}
	\end{equation}
	
	These values are chosen to place the system in a contracting phase approaching a bounce. While not derived from specific physical scenarios, they allow clear demonstration of bouncing dynamics. In realistic cosmological applications, initial conditions would be constrained by late-time observations propagated backwards.
	
	\subsubsection{Numerical Results}
	
\begin{figure}[t]
		\centering
		\includegraphics[width=0.45\textwidth]{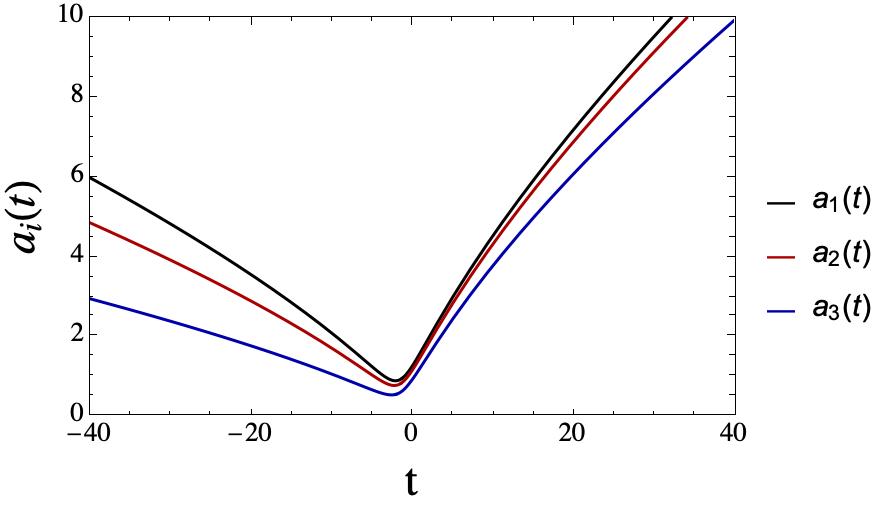}
		\caption{Classical evolution of directional scale factors $a_i(t)$ for Brans-Dicke Bianchi I with $\omega = -5$. All three directions exhibit smooth bounces, with anisotropic structure evident in different bounce amplitudes. Initial conditions are given in Eq.~\eqref{eq:initial_conditions_generic}.}
		\label{fig:classical_generic_scales}
	\end{figure}

	Figure~\ref{fig:classical_generic_scales} shows the evolution of scale factors $a_i(t)$ obtained by numerical integration of Eqs.~\eqref{eq:eom_generic}. All three directional scale factors exhibit smooth bounces near $t \approx 0$, with minimum values $a_i^{\text{min}} \sim (0.8$-$1.1$) depending on direction. The anisotropy is evident: different directions have different bounce times and minimum scales.
	
	\begin{figure}[t]
		\centering
		\includegraphics[width=0.42\textwidth]{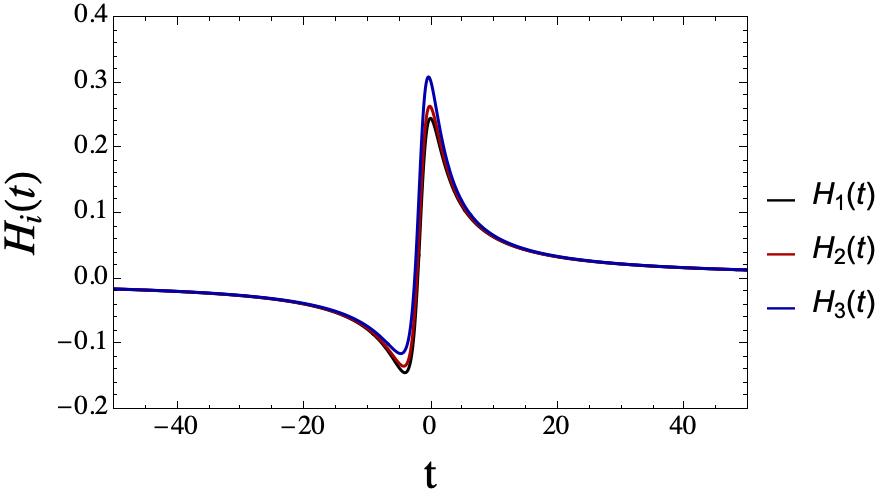}
		\caption{Directional Hubble parameters $H_i(t)$ for the evolution in Fig.~\ref{fig:classical_generic_scales}. Negative values indicate contraction; positive values indicate expansion. Zero crossings mark the bounce points, which occur at slightly different times for different directions due to anisotropy.}
		\label{fig:classical_generic_hubble}
	\end{figure}
	
	Figure~\ref{fig:classical_generic_hubble} displays the corresponding directional Hubble parameters $H_i(t)$ computed from Eq.~\eqref{eq:hubble_directional}. Each $H_i$ transitions from negative (contraction) through zero (bounce) to positive (expansion). The asymptotic behavior is $H_i \to 0$ as $t \to \pm\infty$, indicating approach to Minkowski space at early and late times in this configuration.
	
\begin{figure}[t]
		\centering
		\includegraphics[width=0.4\textwidth]{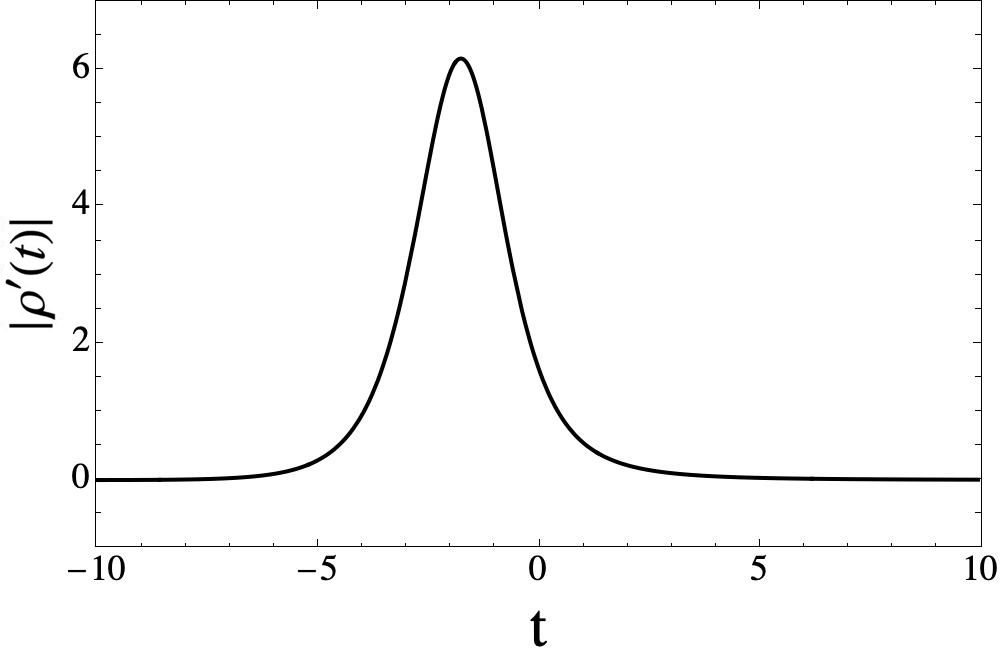}
		\caption{Energy density for $\omega = -5$ case, computed from Eq.~\eqref{eq:energy_density}. The density remains finite at the bounce, reaching $\rho_{\text{max}} \approx 6$. This demonstrates classical energy singularity avoidance in BD theory with $\omega < -3/2$.}
		\label{fig:classical_generic_rho}
	\end{figure}

	The energy density $\rho(t)$ (Fig.~\ref{fig:classical_generic_rho}) remains finite throughout evolution, reaching a maximum $\rho_{\text{max}} \approx 6$ (in Planck units with our normalization) at the bounce. This finiteness is a key feature: despite the classical bounce, no energy singularity occurs.
	
\begin{figure}[t]
		\centering
		\includegraphics[width=0.43\textwidth]{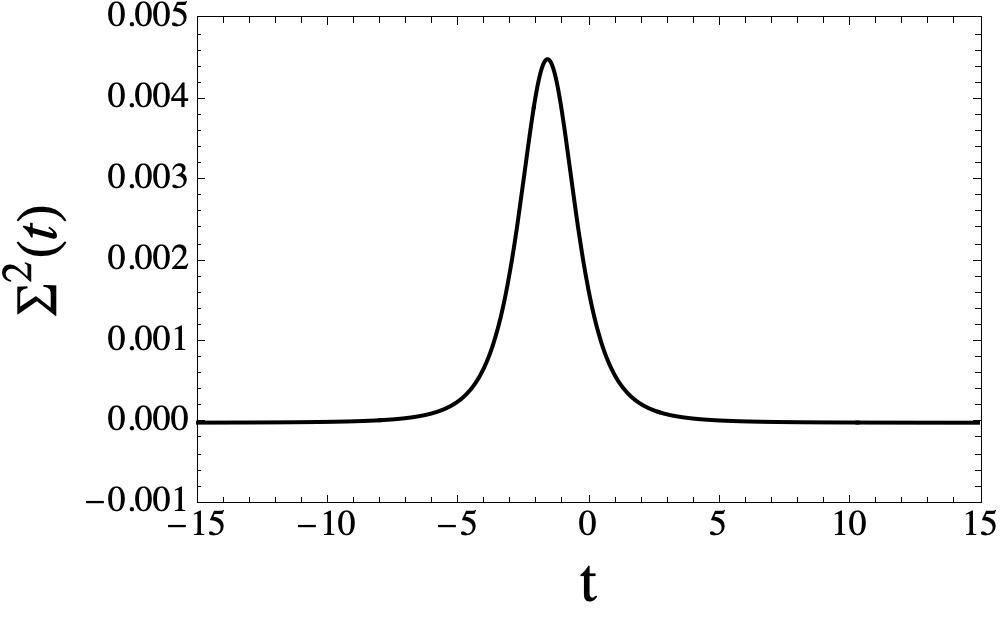}
		\caption{Shear $\Sigma^{2}(t)$ quantifying anisotropy for $\omega = -5$. The shear peaks near the bounce where directional expansion rates differ most significantly, then decays as the universe isotropizes during expansion.}
		\label{fig:classical_generic_shear}
	\end{figure}

	The shear $\Sigma^2(t)$ in Figure~\ref{fig:classical_generic_shear} measures anisotropy. It peaks near the bounce and decays as the universe expands, indicating isotropization at late times. 
	
	\subsection{Classical Dynamics: Case $\omega = -3/2$}
	\label{subsec:classical_conformal}
	
	For the conformally invariant case, Hamilton's equations derived from~\eqref{eq:hamiltonian_conformal} yield (setting $N = \gamma = 1$):
	\begin{eqnarray}
		\label{eq:eom_conformal}
		\dot{c}_1 &= -\frac{1}{2\phi\sqrt{p}} \left( c_1 c_2 p_2 + c_1 c_3 p_3 - \frac{c_2 c_3 p_2 p_3}{p_1} \right) + \lambda c_1, \nonumber \\
		\dot{c}_2 &= -\frac{1}{2\phi\sqrt{p}} \left( c_1 c_2 p_1 + c_2 c_3 p_3 - \frac{c_1 c_3 p_1 p_3}{p_2} \right) + \lambda c_2, \nonumber \\
		\dot{c}_3 &= -\frac{1}{2\phi\sqrt{p}} \left( c_1 c_3 p_1 + c_2 c_3 p_2 - \frac{c_1 c_2 p_1 p_2}{p_3} \right) + \lambda c_3, \nonumber \\
		\dot{p}_1 &= \frac{\sqrt{p}}{\phi} (c_3 p_2 + c_2 p_3) - \lambda p_1, \nonumber \\
		\dot{p}_2 &= \frac{\sqrt{p}}{\phi} (c_3 p_1 + c_1 p_3) - \lambda p_2, \nonumber \\
		\dot{p}_3 &= \frac{\sqrt{p}}{\phi} (c_2 p_1 + c_1 p_2) - \lambda p_3, \nonumber \\
		\dot{\phi} &= -\lambda \phi, \nonumber \\
		\dot{p}\phi &= -\frac{S}{\phi^2 \sqrt{p}} + \lambda p\phi.
	\end{eqnarray}
	The Lagrange multiplier $\lambda$ enters linearly in each equation. It is not a dynamical variable but a constraint multiplier whose value must be determined consistently. It affects the time parametrization but not physical observables expressible as ratios of evolving quantities.
	
	To understand its role, consider the evolution of $\phi$:
	\begin{equation} \label{phievol}
		\phi(t) = \phi(0) e^{-\lambda t}.
	\end{equation}
	Different values of $\lambda$ correspond to different time 
	parametrizations of the same physical spacetime~\cite{lin2023conformally,
		dirac2013lectures}. To make this precise, we use $\phi$ as a 
	relational (internal) clock. 
	From Eq.~\eqref{phievol}, the coordinate time corresponding to 
	a given value of $\phi$ is
	\begin{equation}
		t(\phi;\lambda) = -\frac{1}{\lambda}\ln\frac{\phi}{\phi(0)},
		\label{eq:tphi}
	\end{equation}
	so a rescaling $\lambda \to \mu\lambda$ is equivalent to a 
	coordinate time rescaling $t \to t/\mu$. 
	For numerical studies we will choose $\lambda = 1$ as a reference value.
	\subsubsection{De Sitter Asymptotics}
	An interesting feature of the $\omega$ = -3/2 case is its asymptotic de Sitter behavior~\cite{hrycyna2014dynamics}. From Eq.~\eqref{eq:eom_conformal}, one can show that as $t \to \pm\infty$, each $H_i$ approaches a constant value $H_i \to H_\infty$, implying exponential expansion (or contraction) $a_i \sim e^{H_\infty t}$.
	
	This is related to conformal invariance: the theory admits solutions asymptotically approaching de Sitter space, which is conformally flat. Such behavior has been explored as a mechanism for early universe inflation~\cite{hrycyna2014dynamics} and late-time acceleration~\cite{oikonomou2022effects}.
	\subsubsection{Numerical Results}
	We evolve Eqs.~\eqref{eq:eom_conformal} with initial conditions
	\begin{eqnarray}
		%\begin{aligned}
		c_1(0) = 1.1,\ & p_1(0) = 1.0, \nonumber \\
		c_2(0) = 1.2,\ & p_2(0) = 2.0, \nonumber \\
		c_3(0) = 1.3, \ & p_3(0) = 3.0, \nonumber \\
		\phi(0) = 10, \ & p_\phi(0) = 1.0,
		%\end{aligned}
		\label{eq:initial_conditions_conformal}
	\end{eqnarray}
	and $\lambda = 1$.
	
	While physical observables are $\lambda$-independent when expressed in terms of the relational  $\phi$, we present numerical results as functions of coordinate time $t$ with the gauge choice $\lambda = 1$. This is a valid gauge fixing equivalent to selecting a specific parametrization of the physical trajectory, and it allows direct visual comparison between classical and effective quantum evolutions — the central objective of Sec.~\ref{sec:numerical}.
	
\begin{figure}[t]
		\centering
		\includegraphics[width=0.45\textwidth]{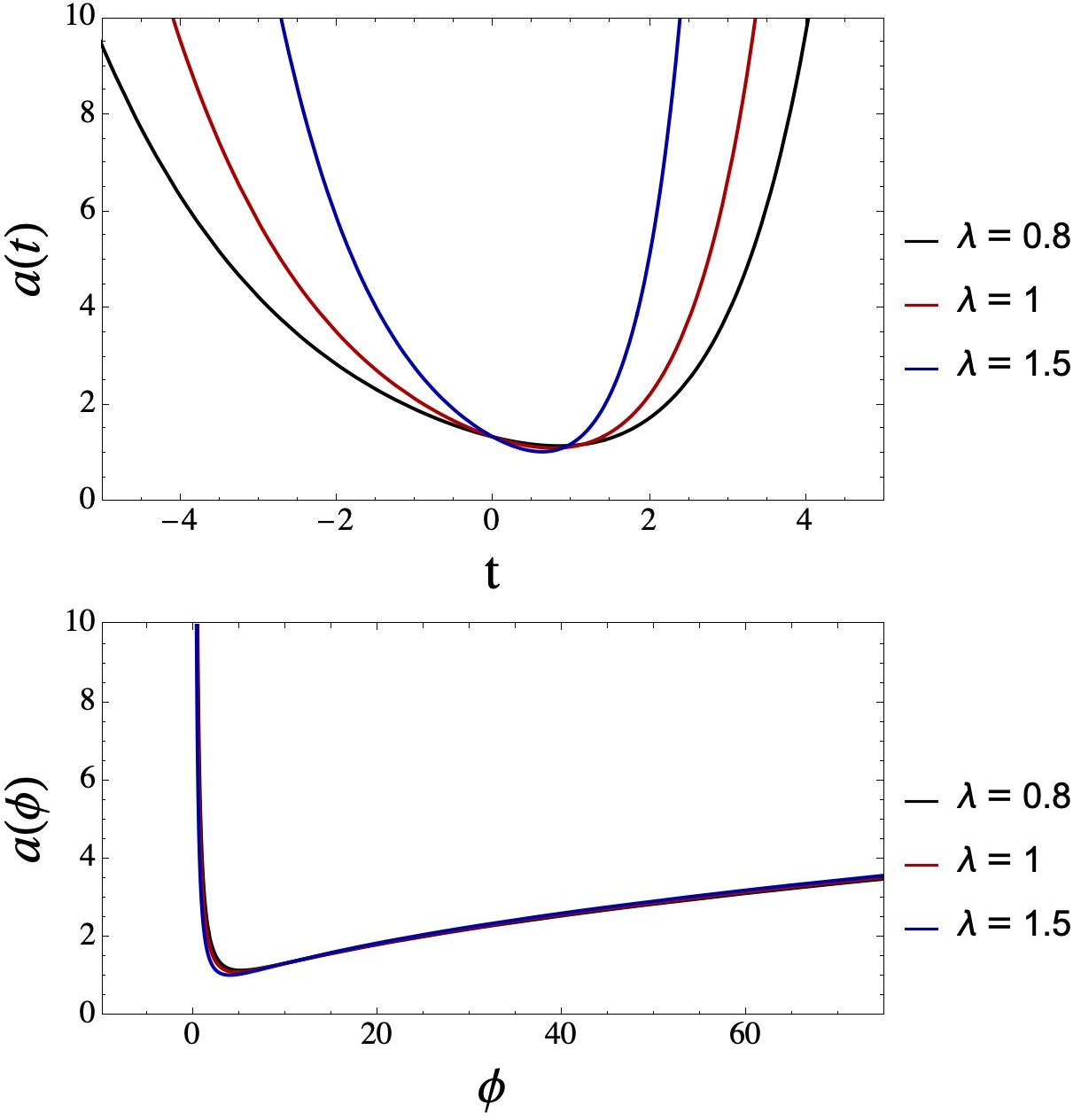}
		\caption{Classical evolution of the mean scale factors $a(t)$ and $a(\phi)$ for $\omega = -3/2$ and different values of $\lambda$. The top panel shows $a(t)$ for different $\lambda$: the apparent $\lambda$-dependence here is a coordinate effect, since different values of $\lambda$ correspond to different time parametrizations $t \to t/\mu$ (see Eq.~\eqref{eq:tphi}). The bottom panel uses the scalar field $\phi$ as a relational (internal) clock, displaying that the $a(\phi)$ evolution is manifestly independent of $\lambda$ and encodes the physical trajectory. Smooth bounces occur in all cases, structurally similar to those found for $\omega = -5$, but with distinct bounce scales and modified asymptotic behavior (top panel). Initial conditions from Eq.~\eqref{eq:initial_conditions_conformal}.}
		\label{fig:classical_conformal_scales}
	\end{figure}

	Figure~\ref{fig:classical_conformal_scales} shows the mean scale factor $a(t)$ evolution for different values of $\lambda$ (top panel). As in the $\omega =-5$ case, smooth bounces occur, but now with different quantitative features due to the modified dynamics. The top panel displays a spurious $\lambda$-dependence that is an artifact of coordinate time; as discussed in Sec.~\ref{subsec:classical_conformal}, different $\lambda$ values correspond to rescaled coordinate times $t\to t/\mu$. The figure also shows the independence of the scale factor on $\lambda$ when $\phi$ is used as internal time (bottom panel), which represents the physical result.
	
\begin{figure}[t]
		\centering
		\includegraphics[width=0.45\textwidth]{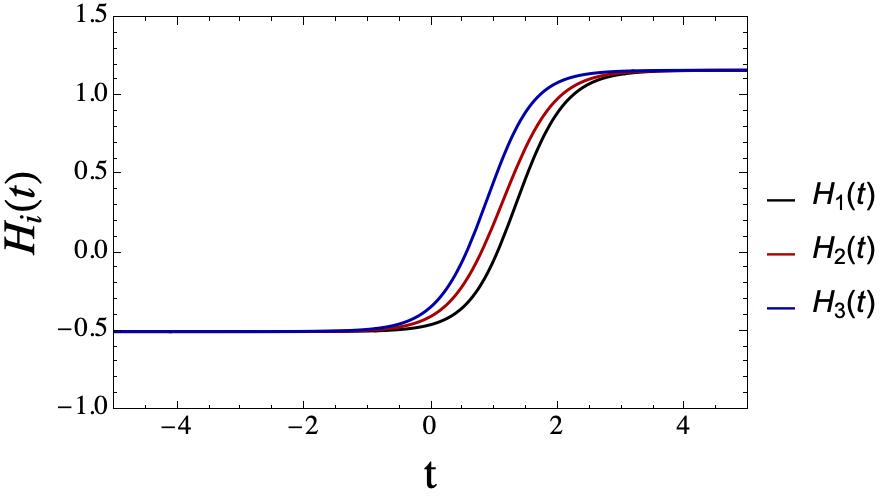}
		\caption{Directional Hubble parameters for $\omega = -3/2$. Unlike the $\omega = -5$ case, here $H_i$ asymptotes to constant values $H_{+\infty} \approx 1.1$, and $H_{-\infty} \approx -0.5$ rather than zero, indicating asymptotic de Sitter contraction (past) and expansion (future). This is a signature of conformal invariance.}
		\label{fig:classical_conformal_hubble}
	\end{figure}

	Figure~\ref{fig:classical_conformal_hubble} displays the Hubble parameters. Instead of asymptoting to zero, each $H_i$ approaches a constant value $H_{+\infty} \approx 1.1$, and $H_{-\infty} \approx -0.5$ (determined by $\lambda$ and initial conditions). This constant Hubble parameter signals de Sitter expansion at late times and de Sitter contraction at early times, with the bounce interpolating between these phases.
	
\begin{figure}[t]
		\centering
		\includegraphics[width=0.45\textwidth]{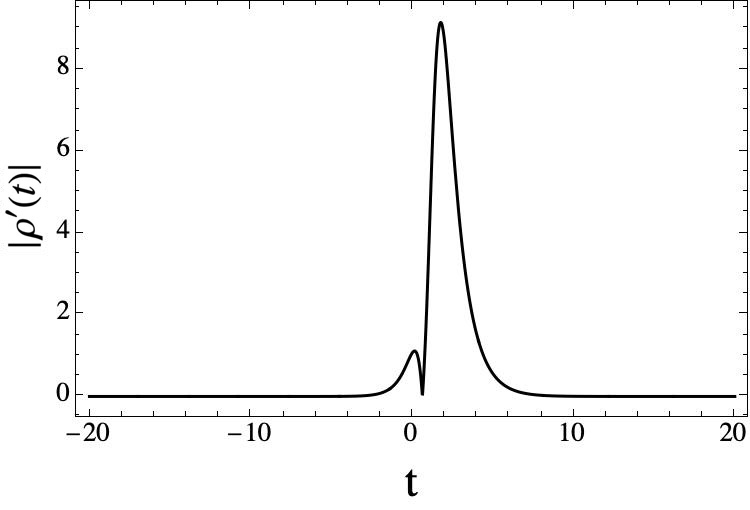}
		\caption{Energy density for $\omega = -3/2$, remaining finite at the bounce with $\rho_{\text{max}} \approx 9$.}
		\label{fig:classical_conformal_rho}
	\end{figure}

\begin{figure}[t]
		\centering
		\includegraphics[width=0.45\textwidth]{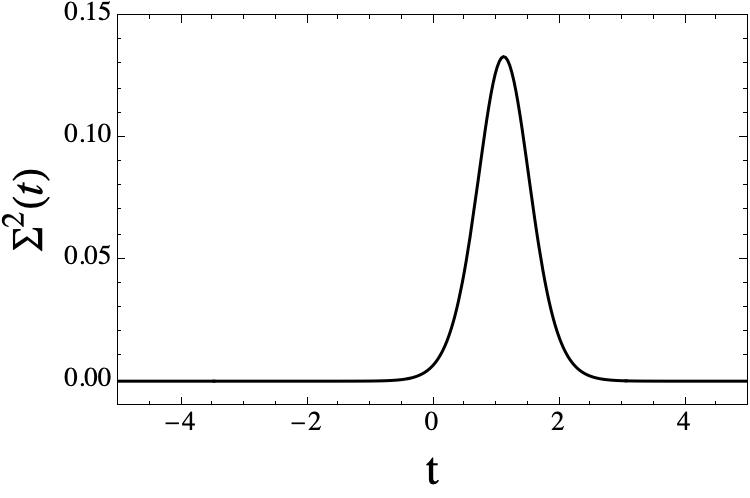}
		\caption{Shear evolution for $\omega = -3/2$, showing peak near bounce and subsequent decay.}
		\label{fig:classical_conformal_shear}
	\end{figure}

	Energy density and shear (Figs. \ref{fig:classical_conformal_rho} and \ref{fig:classical_conformal_shear}) exhibit qualitatively similar behavior to the $\omega = -5$ case: finite $\rho$ at the bounce and shear peaking near the bounce then decaying.
	
	
	\subsection{Summary of Classical Dynamics}
	Both cases $\omega = -5$ and $\omega = -3/2$ exhibit:
	\begin{itemize}
		\item Smooth bouncing solutions with finite energy density;
		\item Anisotropic evolution (different bounce times and amplitudes for different directions);
		\item Classical singularity avoidance via the BD scalar field mechanism.
	\end{itemize}
	
	Key differences:
	\begin{itemize}
		\item Asymptotic behavior: $H_i \to 0$ for $\omega = -5$; $H_i \to H_\infty \neq 0$
		for $\omega = -3/2$;
		\item Constraint structure: Single Hamiltonian constraint vs. Hamiltonian plus conformal constraint
		\item Time parametrization: Standard for $\omega < -3/2$; $\lambda$-dependent for $\omega = -3/2$.
	\end{itemize}
	
	In the following sections, we investigate how quantum effects modify these classical bouncing scenarios.
	%%%%%%%%%%%%%%%%%%%%%%%%%%%%%%%%%%
	%
	\section{Effective Quantum Dynamics}
	\label{sec:effective}
	We now develop the effective quantum formalism for the Brans-Dicke Bianchi I model. After reviewing the general effective approach (Sec.\ref{subsec:effective_general}), we construct the second-order effective Hamiltonian (Sec.\ref{subsec:effective_hamiltonian}) and discuss its regime of validity (Sec.~\ref{subsec:validity}).
	\subsection{General Effective Formalism}
	\label{subsec:effective_general}
	
	We provide now an effective quantum analysis of the model at hand by working directly with expectation values and quantum moments.
	
	\subsubsection{Quantum Moments and Extended Phase Space}
	Consider a quantum system with canonical variables $(\hat{q}_i, \hat{p}_i)$
	satisfying $[\hat{q}_i, \hat{p}_j] = i\hbar \delta_{ij}$. We define expectation values
	\begin{equation}
		\langle \hat{q}_i \rangle \equiv q_i, \quad \langle \hat{p}_i \rangle \equiv p_i,
	\end{equation}
	and quantum moments (using Weyl ordering for symmetrization):
	\begin{equation}
		\Delta(q_i^a p_j^b) \equiv \langle (\hat{q}_i - q_i)^a (\hat{p}_j - p_j)^b \rangle_{\text{Weyl}}.
		\label{eq:quantum_moment}
	\end{equation}
	
	With $a + b = 2$. The correlations and moments:
	\begin{subequations}
		\begin{align}
			\Delta(q_i^2) &= \langle (\hat{q}_i - q_i)^2 \rangle = (\Delta q_i)^2, \\
			\Delta(p_i^2) &= \langle (\hat{p}_i - p_i)^2 \rangle = (\Delta p_i)^2, \\
			\Delta(q_i p_i) &= \frac{1}{2}\langle \{\hat{q}_i - q_i, \hat{p}_i - p_i\} \rangle .
		\end{align}
	\end{subequations}
	
	For $i \neq j$, $\Delta(q_i p_j)$, $\Delta(q_i q_j)$, and $\Delta(p_i p_j)$ are the cross-correlations between different degrees of freedom.
	
	The quantum moments satisfy Heisenberg uncertainty relations. For a single degree of freedom:
	\begin{equation}
		\Delta(q^2) \Delta(p^2) - \Delta(qp)^2 \geq \frac{\hbar^2}{4}.
		\label{eq:heisenberg_single}
	\end{equation}
	For multiple degrees of freedom, more complex uncertainty relations exist, involving cross-correlations~\cite{bojowald2012quantum}.
	\subsubsection{Effective Hamiltonian and Equations of Motion}
	The effective Hamiltonian is defined as the expectation value of the quantum Hamiltonian operator:
	\begin{equation}
		H_{\text{eff}} = \langle \hat{H} \rangle.
	\end{equation}
	Expanding in powers of quantum moments:
	\begin{equation}
		H_{\text{eff}} = H_{\text{cl}}(q, p) + \sum_{n=2}^\infty H^{(n)}(\Delta(q^a p^b)),
		\label{eq:heff_expansion}
	\end{equation}
	where $H_{\text{cl}}$ is the classical Hamiltonian and $H^{(n)}$ contains contributions from moments of order $n$.
	
	Formally, this expansion can be derived via Taylor expansion~\cite{bojowald2006effective}:
	\begin{equation}
		H_{\text{eff}} = \sum_{a,b} \frac{1}{a! b!} \frac{\partial^{a+b} H_{\text{cl}}}{\partial q^a \partial p^b} \bigg|_{q,p} \Delta(q^a p^b).
		\label{eq:taylor_expansion}
	\end{equation}
	
	The equations of motion follow from an effective Poisson bracket structure~\cite{bojowald2006effective}:
	\begin{eqnarray}
		\dot{f} &=& \{ f, H_{\text{eff}} \}, \nonumber \\
		\{ \langle \hat{f} \rangle, \langle \hat{g} \rangle \} &=& \frac{1}{i \hbar} \langle [\hat{f}, \hat{g}] \rangle .
		\label{eq:effective_poisson}
	\end{eqnarray}
	The first one applies for classical variables, and the second one for momenta.
	
	For quantum moments, the bracket algebra can be derived from commutation relations. For example:
	\begin{eqnarray}
		\{ \Delta(q^2), \Delta(p^2)\} &= 4 \Delta(qp), \nonumber \\
		\{ \Delta(q^2), \Delta(qp) \} &= 2 \Delta(q^2), \nonumber \\
		\{ \Delta(p^2), \Delta(qp) \} &= -2 \Delta(p^2).
	\end{eqnarray}
	\subsubsection{Truncation and Hierarchy}
	The expansion~\eqref{eq:heff_expansion} is, in general, infinite, so, for practical calculations, we truncate at finite order. Since moments scale as $\Delta(q^a p^b) \sim \hbar^{(a+b)/2}$, the expansion is effectively in powers of $\hbar$. Truncation at second order yields the following semiclassical approximation:
	\begin{equation}
		H_{\text{eff}} \approx H_{\text{cl}} + H^{(2)}(\Delta(q^2), \Delta(p^2), \Delta(qp), \ldots) + \mathcal{O}(\hbar^{3/2}).
		\label{eq
			:second_order_truncation}
	\end{equation}
	This truncation is justified when the system is in a semiclassical regime where quantum fluctuations are small compared to expectation values $\sqrt{\Delta(q^2)} \ll |q|$, and higher-order moments remain dynamically suppressed: $\Delta(q^3 p^3)| \ll |\Delta(q^2 p^2)|$.
	
	\subsection{Effective Hamiltonian for Brans-Dicke Bianchi I}
	\label{subsec:effective_hamiltonian}
	
	\subsubsection{Phase Space and Quantum Variables}
	Our system has $k = 4$ canonical pairs: $(c_i, p_i)$ for $i = 1, 2, 3$ and $(\phi, p_\phi)$. At second order, the relevant quantum variables are:
	\begin{itemize}
		\item Expectation values (8): $c_1, c_2, c_3, p_1, p_2, p_3, \phi, p_\phi$
		\item Diagonal moments (8): $\Delta(c_i^2), \Delta(p_i^2)$ for $i = 1, 2, 3$; $\Delta(\phi^2),\Delta(p_\phi^2)$
		\item Covariances (4): $\Delta(c_i p_i)$ for $i = 1, 2, 3$; $\Delta(\phi p_\phi)$
		\item Cross-correlations (28): $\Delta(c_i c_j),\Delta(p_i p_j), \Delta(c_i p_j)$ for $i \neq j$; $\Delta(c_i \phi), \Delta(c_i p_\phi), \Delta(p_i \phi), \Delta(p_i p_\phi)$.
		%\textcolor{red}{[explicitly: 3 terms of type $\Delta(c_i c_j)$, 3 of type $\Delta(p_i p_j)$, 6 of type $\Delta(c_i p_j)$ with $i\neq j$, 3 of type $\Delta(c_i\phi)$, 3 of type $\Delta(c_i p_\phi)$, 3 of type $\Delta(p_i\phi)$, 3 of type $\Delta(p_i p_\phi)$; the remaining 4 covariances $\Delta(q_a p_a)$ are listed separately above: total $3+3+6+3+3+3+3+4=28$ off-diagonal second-order moments]}
	\end{itemize}
	
	This yields an extended phase space of dimension 48.
	
	\subsubsection{Second-Order Effective Hamiltonian: $\omega < -3/2$.}
	
	Applying the Taylor expansion~\eqref{eq:taylor_expansion} to the classical Hamiltonian~\eqref{eq:hamiltonian_generic} and computing derivatives up to second order, we obtain (setting $N = \gamma = 1$):
	\begin{equation}
		\begin{split}
			H_{\text{eff}}^{(\omega)} = & H_{\text{BD}} + \frac{1}{\sqrt{p}} \Bigg[
			\sum_{i=1}^3 \left( \frac{p_i^2 \zeta}{\phi} \Delta(c_i^2) + \frac{1}{2\phi} \mathcal{A}_i \Delta(p_i^2) \right) \\
			& + \frac{\zeta \phi}{\phi} \Delta(p_\phi^2) + \frac{1}{2\phi^2} \mathcal{B} \Delta(\phi^2) + 2\zeta \Delta(\phi p_\phi) \\
			& + \sum_{i=1}^3 \left( \mathcal{C}_i \Delta(c_i p_i) + \frac{2p_i \zeta}{\sqrt{p}} \Delta(c_i p_\phi) + \mathcal{D}_i \Delta(c_i \phi) \right) \\
			& + \sum_{i < j} \left( \mathcal{E}_{ij} \Delta(c_i c_j) + \mathcal{F}_{ij} \Delta(c_i p_j) + \mathcal{G}_{ij} \Delta(p_i p_j) \right) \\
			& + \sum_{i=1}^3 \left( \mathcal{J}_i \Delta(p_i \phi) + \mathcal{K}_i \Delta(p_i p_\phi) \right) \Bigg],
		\end{split}
		\label{eq:heff_generic_compact}
	\end{equation}
	where $H_{\text{BD}}$ is given by Eq.~\eqref{eq:hamiltonian_generic} evaluated at expectation values, and the coefficient functions $\mathcal{A}_i, \mathcal{B}, \mathcal{C}_i$, etc., are second derivatives of $H_{\text{BD}}$ evaluated at $(c_i, p_i, \phi, p_\phi)$. As illustrative examples, the leading coefficients are:
	\begin{subequations}
		\begin{align}
			\mathcal{A}_i &= \frac{\partial^2 H_{\text{BD}}}{\partial p_i^2} = \frac{1}{\phi\sqrt{p}}\left(\frac{S - \zeta T^2}{4p_i^2} - \frac{\zeta T}{p}\right), \\
			\mathcal{B} &= \frac{\partial^2 H_{\text{BD}}}{\partial \phi^2} = -\frac{2}{\phi^3\sqrt{p}}(S - \zeta T^2) + \frac{2\zeta\gamma^2 p_\phi^2}{\phi\sqrt{p}}, \\
			\mathcal{C}_i &= \frac{\partial^2 H_{\text{BD}}}{\partial c_i \partial p_i} = -\frac{1}{\phi\sqrt{p}}\left(c_j p_j + c_k p_k - 2\zeta T\right)\bigg|_{j,k\neq i}.
		\end{align}
	\end{subequations}
	The complete set of coefficient functions is implemented numerically; see Appendix~\ref{app:eom} for the structure of the resulting equations of motion.
	
	\subsubsection{Second-Order Effective Hamiltonian: $\omega = -3/2$.}
	
	For the conformally invariant case, we expand~\eqref{eq:hamiltonian_conformal}:
	\begin{eqnarray}
		H_{\text{eff}}^{(c)} &=& H_{\text{BD}}^c - \lambda \Delta(\phi p_\phi) \nonumber \\
		&& + \frac{1}{\phi^2 \sqrt{p}} \Bigg[
		\sum_{i=1}^3 \left( \mathcal{A}_i^c \Delta(p_i^2) + \phi \mathcal{A}_i^c \Delta(c_i p_i) \right) \nonumber \\
		&& -\frac{S}{\phi} \Delta(\phi^2) + \sum_{i=1}^3 \left( \mathcal{D}_i^c \Delta(c_i \phi) + \mathcal{J}_i^c \Delta(p_i \phi) \right) \nonumber \\
		&& + \sum_{i < j} \left( \mathcal{E}_{ij}^c \Delta(c_i c_j) + \mathcal{F}_{ij}^c \Delta(c_i p_j) + \mathcal{G}_{ij}^c \Delta(p_i p_j) \right) \nonumber \\
		&& - \phi \sum_{i=1}^3 \Delta(c_i c_j)|_{j \neq i} \Bigg].
		\label{eq:heff_conformal_compact}
	\end{eqnarray}
	
	\subsection{Validity and Limitations of the Effective Approach}
	\label{subsec:validity}
	\subsubsection{Semiclassicality Conditions}
	For the second-order truncation to be reliable, quantum fluctuations must remain small compared to expectation values. Define dimensionless ratios:
	\begin{equation}
		r_{c_i} \equiv \frac{\sqrt{\Delta(c_i^2)}}{|c_i|}, \quad
		r_{p_i} \equiv \frac{\sqrt{\Delta(p_i^2)}}{|p_i|}.
		\label{eq:semiclassicality_ratios}
	\end{equation}
	Semiclassicality requires $r_{c_i}, r_{p_i}$ to be small far from the bounce. We verify this condition a posteriori in our numerical solutions (Sec.~\ref{subsec:numerical_generic}).
	Additionally, for the Taylor expansion to converge, higher-order moments must be suppressed. A sufficient condition is that the quantum state is nearly Gaussian, since Gaussian states satisfy $\Delta(q^{2n} p^{2m}) = (2n-1)!! (2m-1)!! [\Delta(q^2)]^n [\Delta(p^2)]^m$ (for uncorrelated $q, p$), yielding automatic suppression of higher orders in the semiclassical limit.
	
	\subsubsection{Quantum Gravity Scale and Breakdown}
	The effective approach describes quantum corrections to classical dynamics but does not capture full quantum gravity effects such as spacetime superpositions or topology change. It is expected to break down when:
	\begin{enumerate}
		\item Curvature approaches Planck scale: $R \sim \ell_{\text{Pl}}^{-2}$.
		\item Volume approaches Planck scale: $V \sim \ell_{\text{Pl}}^3$.
		\item Quantum fluctuations become of order unity: $r_{c_i}, r_{p_i} \sim 1$.
	\end{enumerate}
	
	For our Brans-Dicke model, the effective gravitational constant is $G_{\text{eff}} = 1/\phi$. The Planck length is therefore $\ell_{\text{Pl}} \sim \phi^{-1/2}$. Near the bounce, with $\phi(0) = 10$ (in our units), we have $\ell_{\text{Pl}} \sim 0.3$. The minimum scale factor values $a_i^{\text{min}} \sim 1$ correspond to volumes $V \sim 1$, which is $\sim 10^{1.5}$
	times the Planck volume. Thus, we are on the edge of the semiclassical regime, where quantum corrections are significant but the effective approach is still valid.
	
	As a consistency check, we will monitor the ratios~\eqref{eq:semiclassicality_ratios} throughout evolution, satisfaction of Heisenberg uncertainty relations~\eqref{eq:heisenberg_single} and the energy density magnitude.
	
	\subsubsection{Comparison with Loop Quantum Cosmology}
	A natural question is how our effective approach compares to full loop quantization. Loop quantum cosmology for Brans-Dicke Bianchi I has been studied in Refs.~\cite{zhang2012loop,sharma2025quantum}. Among the most important differences we can mention the following
	\begin{itemize}
		\item LQC quantizes the connection via holonomies, introducing a fundamental discreteness scale $\delta \sim \sqrt{\Delta(a_i^2)}$. Bounces occur when $\delta \sim \ell_{\text{Pl}}$.
		\item Our effective approach treats quantum fluctuations as continuous but small. The bounce scale is set by the quantum state width $\sigma_\chi$ (see Sec.~\ref{subsec:initial_conditions}).
	\end{itemize}
	For highly squeezed states ($\sigma_\chi \to 0$), our effective approach should approximate LQC results. For broader states, deviations are expected. 
	
	\subsection{Initial Conditions for Quantum Moments}
	\label{subsec:initial_conditions}
	To evolve the effective system, we must specify initial values for all 48 phase space variables. For expectation values, we use the same values as in the classical cases (Eqs.\eqref{eq:initial_conditions_generic},\eqref{eq:initial_conditions_conformal}).
	For quantum moments, we assume initially uncorrelated Gaussian states. A Gaussian state with width $\sigma_\chi$ has wavefunction
	\begin{equation}
		\psi(\chi) = (\pi \sigma_\chi^2)^{-1/4} \exp\left[ -\frac{(\chi - \chi_0)^2}{2\sigma_\chi^2} + \frac{i p_0 \chi}{\hbar} \right],
		\label{eq
			:gaussian_state}
	\end{equation}
	yielding moments~\cite{bojowald2012quantum}
	\begin{eqnarray}
		\Delta(\chi^{2n})& =& \frac{(2n-1)!!}{2^n} \sigma_\chi^{2n}, \nonumber \\
		\Delta(p_\chi^{2m}) &=& \frac{(2m-1)!!}{2^m} \frac{\hbar^{2m}}{\sigma_\chi^{2m}}, \nonumber \\
		\Delta(\chi^a p_\chi^b) &=& 0, \quad  a+b \in 2 \mathbb{N}+1 .
		\label{eq:gaussian_moments}
	\end{eqnarray}
	For our multi-variable system, assuming factorized Gaussians:
	\begin{eqnarray}
		\label{eq:initial_moments}
		\Delta(c_i^2)(0) = \frac{\sigma_\chi^2}{2}, & \Delta(p_i^2)(0) = \frac{\hbar^2}{2\sigma_\chi^2}, & \Delta(c_i p_i)(0) = 0, \nonumber \\
		\Delta(\phi^2)(0) = \frac{\sigma_\chi^2}{2}, & \Delta(p_\phi^2)(0) = \frac{\hbar^2}{2\sigma_\chi^2}, & \Delta(\phi p_\phi)(0) = 0, \nonumber \\
		\Delta(c_i c_j)(0) = 0, & \Delta(p_i p_j)(0) = 0 & \text{for } i \neq j, \nonumber \\
		\Delta(c_i p_j)(0) = 0, & \Delta(c_i \phi)(0) = 0, & \Delta(c_i p_\phi)(0) = 0, \nonumber \\
		\Delta(p_i \phi)(0) = 0, & \Delta(p_i p_\phi)(0) = 0 & \text{for all } i .
	\end{eqnarray}
	
	The parameter $\sigma_\chi$ controls the initial quantum state width. Smaller $\sigma_\chi$ yields narrower position spread and larger momentum spread, approaching a classical peaked state. Larger $\sigma_\chi$ increases quantum fluctuations.
	We set $\hbar = 1$ (Planck units) and vary $\sigma_\chi$.
	
	In loop quantum cosmology, the bounce scale is set by the area gap $\Delta_{\text{LQC}} \sim \sqrt{\Delta(a_i^2)}$ which is operator-eigenvalue-dependent. In our effective approach, $\sigma_\chi$ plays an analogous role, it sets the scale of quantum fluctuations and thereby influences the bounce scale and quantum correction magnitudes.  Here, we treat it as a free parameter to explore sensitivity.
	%%%%%%%%%%%%%%%%%%%%%%%%%%%%%%%%%%%%%
	
	\section{Numerical Evolution}
	\label{sec:numerical}
	
	We now present numerical solutions of the effective evolution equations for both coupling constant regimes. Our primary focus is demonstrating the crucial role of cross-correlation terms and quantifying quantum backreaction effects. We present results for $\omega < -3/2$ (Sec.~\ref{subsec:numerical_generic}) and $\omega = -3/2$ (Sec.~\ref{subsec:numerical_conformal}), and conclude with a discussion of the validity of our numerical analysis (Sec.~\ref{subsec:numerical_methods}).
	
	\subsection{Effective Evolution: Case $\omega < -3/2$}
	\label{subsec:numerical_generic}
	
	\subsubsection{Critical Importance of Cross-Correlations}
	
	We begin by demonstrating that cross-correlation terms are essential for obtaining physically consistent effective dynamics.
	
	Figure~\ref{fig:effective_scales_generic_3d} compares the evolution of the mean scale factor $a(t)$ with (bottom) and without (top) cross-correlation terms. When cross-correlations are neglected (setting all $\Delta(c_i p_j)|_{i \neq j} = 0$, $\Delta(c_i \phi) = 0$, etc., but evolving diagonal moments), the effective dynamics exhibits pathologies, such that, spurious divergences shortly after the bounce and the violation of the Heisenberg uncertainty relations at late times.
	When cross-correlations are included, all pathologies disappear, and physically reasonable evolution emerges.
	
	The physical interpretation of this  is that cross-correlations encode quantum entanglement between different degrees of freedom. In our system, directional scale factors $a_i$ are coupled through the Hamiltonian constraint. Quantum mechanically, this coupling induces correlations between momentum uncertainties $\Delta(p_i p_j)$ and between canonical pairs of different directions $\Delta(c_i p_j)$. Similarly, the BD scalar $\phi$ couples to geometry, inducing $\Delta(c_i \phi)$ and $\Delta(p_i p_\phi)$ correlations.
	
	Neglecting these correlations effectively assumes degrees of freedom remain separable---a factorization assumption that is generically violated in interacting quantum systems. The divergences arise because this incorrect assumption breaks the symplectic structure of the truncated quantum phase space: setting cross-correlations to zero while evolving only diagonal moments is inconsistent with the Poisson algebra of quantum moments, leading to violations of quantum consistency conditions (such as the Heisenberg uncertainty relations).
	
	Previous effective cosmology studies that neglected cross-correlations~\cite{bojowald2007effective} focused on simpler systems (e.g., FLRW with single matter field) where symmetry prevents some correlations from developing. In more complex scenarios like ours, correlations are dynamically generated and cannot be ignored.
	
	\subsubsection{Evolution of Mean Scale factor and Quantum Smoothing}
	
	Once the cross-correlation terms are included, the effective evolution of the mean scale factor shows a less pronounced bounce compared to the classical trajectory in both the contraction and expansion phases (Figure (\ref{fig:effective_scales_generic_3d}) below). As $\sigma_\chi$ increases (representing larger quantum fluctuations), the transition through the bounce becomes more gradual. For small $\sigma_\chi$ (nearly classical), the effective bounce is sharper, and for larger $\sigma_\chi $ the bounce is spread over a wider time interval. 
	\begin{figure}[t]
		\centering
		\includegraphics[width=0.45\textwidth]{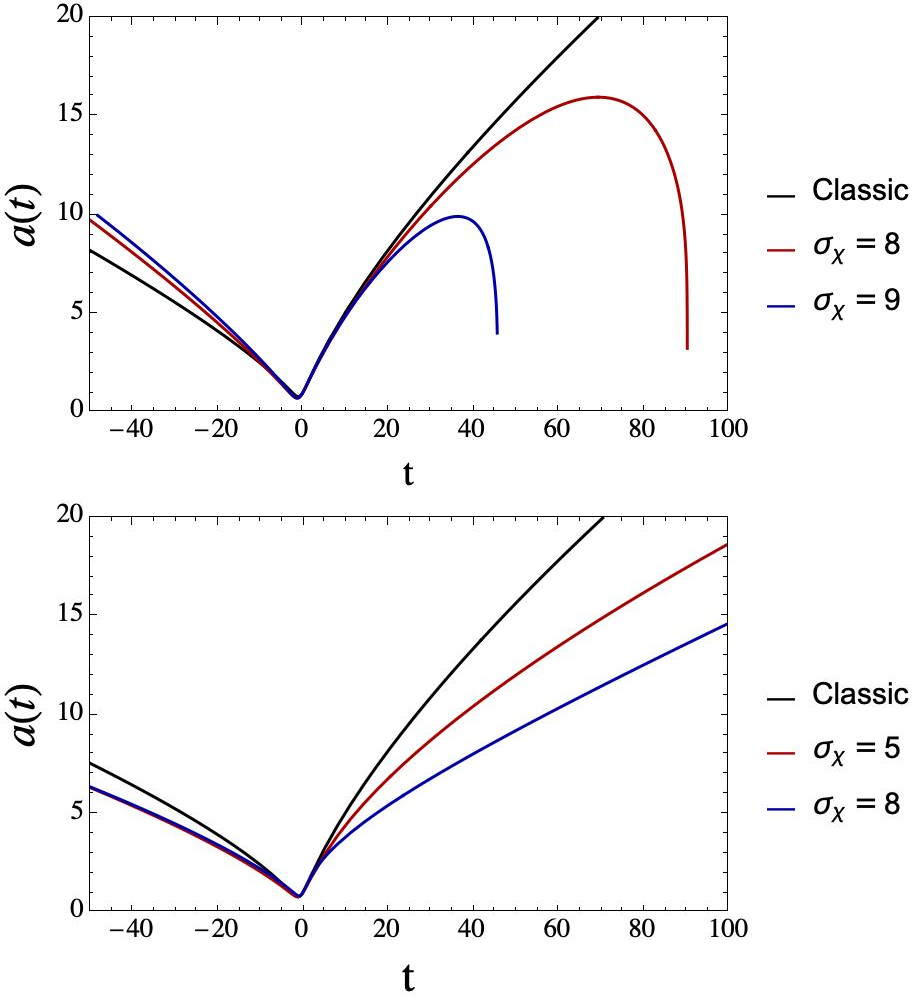}
		\caption{Effective evolution of $a(t)$ for $\omega = -5$ with and without cross-correlation terms included, shown for different quantum state widths $\sigma_\chi$. Top: pathologies appear in the evolution when cross-correlations are not taken into account, causing a recolapse of the scale factor. Bottom: pathologies disappear when cross-correlation terms are included, providing a physically reasonable evolution. As $\sigma_\chi$ increases, quantum effects become more pronounced and the bounce becomes smoother and more gradual.}
		\label{fig:effective_scales_generic_3d}
	\end{figure}
	
	\subsubsection{Hubble Parameters and Post-Bounce Oscillations}
	\label{subsubsec:oscillations}
	
\begin{figure}[t]
		\centering
		\includegraphics[width=0.45\textwidth]{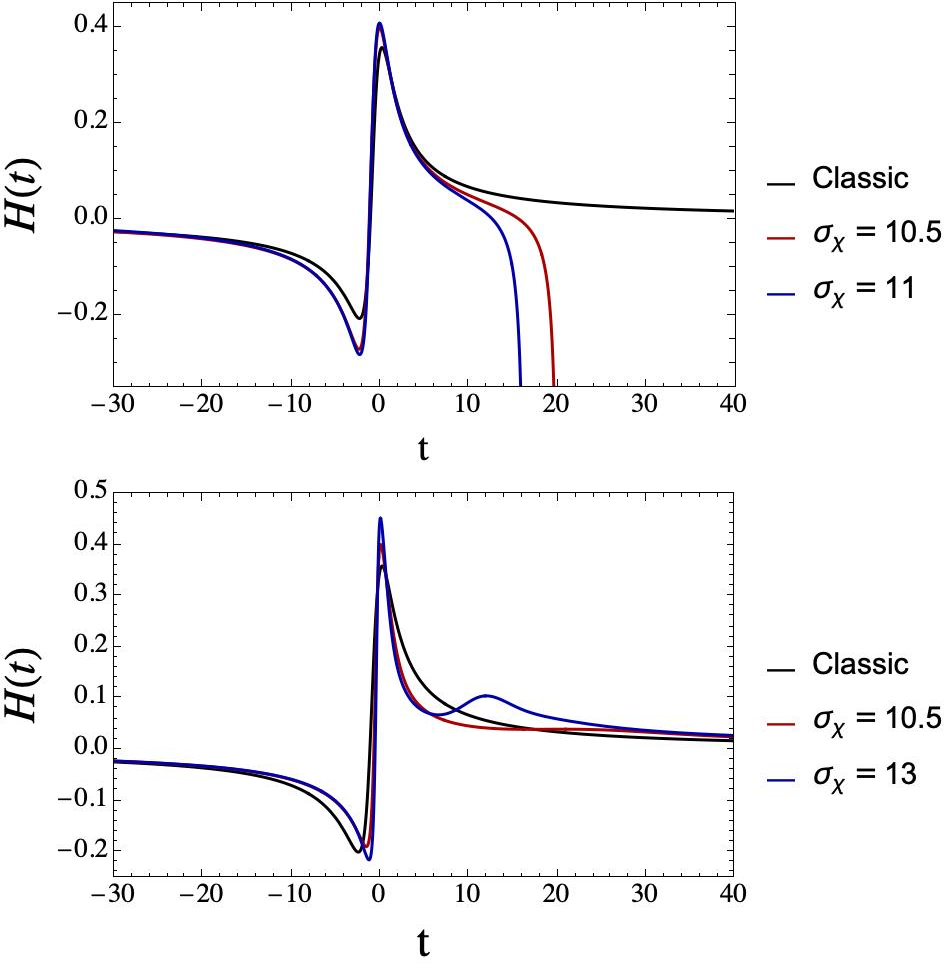}
		\caption{Effective evolution of $H(t)$ for $\omega = -5$ with (bottom) and without (top) cross-correlations. As in figure (\ref{fig:effective_scales_generic_3d}) the cross-correlation terms alleviate the pathologies in the evolution. For small $\sigma_\chi$, evolution closely tracks classical behavior. For larger values of $\sigma_\chi$, characteristic oscillations appear in the post-bounce expansion phase.}
		\label{fig:effective_hubble_generic_3d}
	\end{figure}

	Figure~\ref{fig:effective_hubble_generic_3d} shows the mean directional Hubble parameter $H(t)$ for various $\sigma_\chi$. The classical evolution transitions smoothly from negative (contraction) through zero (bounce) to positive (expansion). As in figure (\ref{fig:effective_scales_generic_3d}), the pathologies in the effective evolution disappear once the cross-correlation terms are included.
	
	For large $\sigma_\chi$, the effective mean Hubble exhibit damped oscillations in the post-bounce regime, rapidly damping to classical behavior
	within a few time units, with initial amplitude $A_{\text{osc}} \sim 0.2$ for $\sigma_\chi = 13$ and an exponential damping.
	
	These oscillations represent quantum remnant effects, transient departures from classical trajectories encoding information about quantum correlations that classical dynamics cannot capture. They arise because near the bounce, quantum moments grow significantly as expectation values become small. As we can notice cross-correlations like $\Delta(c_i p_j)$ and $\Delta(p_i p_j)$ develop complex time dependence and after the bounce, these correlations evolve with characteristic timescales set by the Hamiltonian. Their oscillatory evolution back-reacts on expectation values, producing oscillations in $\langle H_i \rangle$; as the universe expands and quantum effects weaken (increasing volume, decreasing curvature), oscillations damp exponentially.
	
	Mathematically, this can be understood from the structure of equations of motion (Appendix~\ref{app:eom}). The time derivatives of expectation values depend linearly on quantum moments:
	\begin{equation}
		\dot{\langle c_i \rangle} \sim \frac{\partial^2 H}{\partial p_i^2} \Delta(c_i p_i) + \frac{\partial^2 H}{\partial p_i \partial p_j} \Delta(c_i p_j) + \ldots
	\end{equation}
	The moments themselves satisfy coupled oscillator equations with damping terms. This coupled system exhibits damped oscillations characteristic of driven harmonic oscillators. These oscillations only appear when cross-correlations are included. Without them the system lacks the coupled structure necessary for oscillatory back-reaction, further demonstrating their physical necessity. 
	
	\subsubsection{Energy Density Evolution}
	
\begin{figure}[t]
		\centering
		\includegraphics[width=0.45\textwidth]{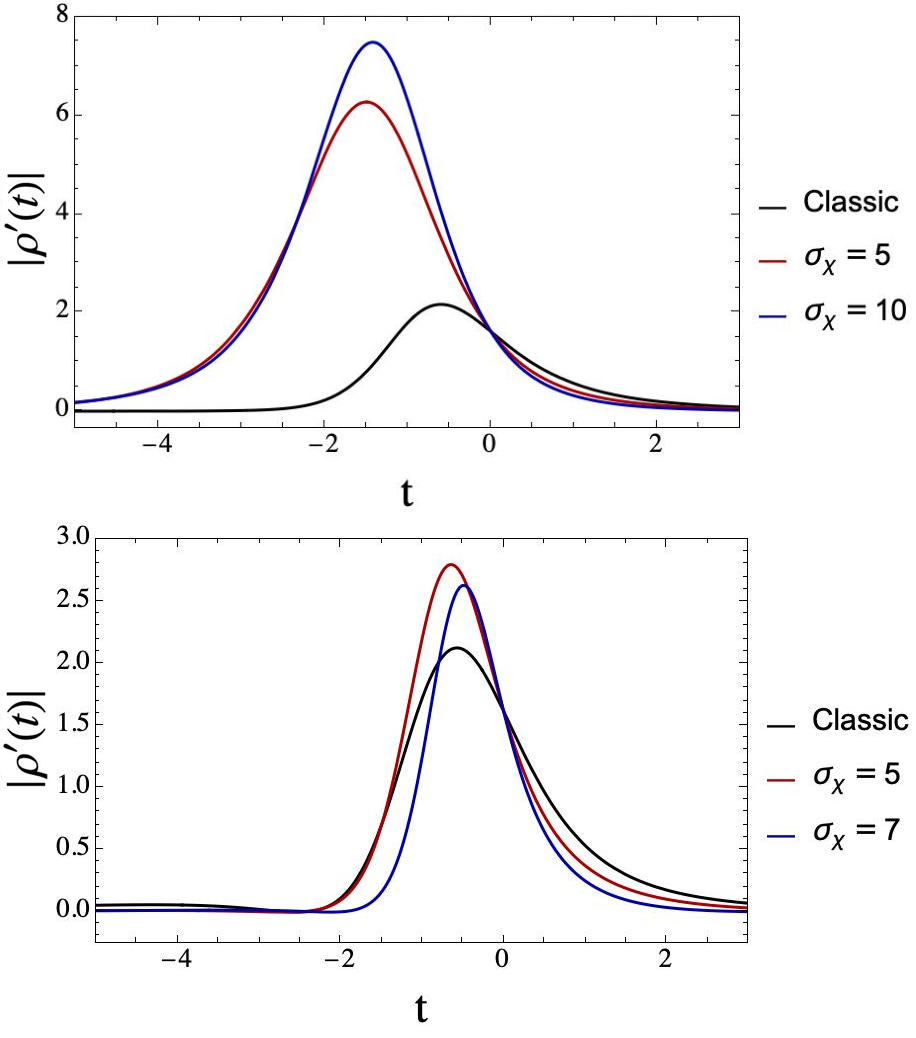}
		\caption{Effective energy density $\rho'(t)$ for $\omega = -5$ 
			with cross-correlations (bottom) and without cross-correlations 
			(top), for various $\sigma_\chi$. Without cross-correlations 
			(top), the energy density develops a spurious unphysical spike 
			near the bounce, an artifact of the inconsistent truncated 
			dynamics. With cross-correlations included (bottom), the peak 
			remains sharply localized near the bounce but is slightly 
			enhanced relative to the classical value. All effective 
			evolutions converge to the classical result at late times.}
		\label{fig:effective_rho_generic_3d}
	\end{figure}

	The effective energy density, computed from 
	Eq.~\eqref{eq:energy_density} with quantum corrections, 
	is shown in Fig.~\ref{fig:effective_rho_generic_3d}.
	
	Key features displayed are:
	\begin{itemize}
		\item \emph{Peak enhancement:} With cross-correlations included, 
		the peak value $\rho'_{\max}$ is slightly increased compared 
		to the classical case. The energy density remains sharply localized 
		near the bounce, qualitatively similar to the classical profile 
		but with a higher amplitude. This indicates that quantum 
		backreaction slightly concentrates energy near the bounce.
		\item \emph{Elimination of unphysical spike:} Without 
		cross-correlations (top panel), the energy density develops 
		a spurious divergent spike near the bounce. Including all 
		cross-correlation terms removes this pathology entirely, 
		yielding a smooth, physically consistent profile.
		\item \emph{Late-time convergence:} All effective evolutions 
		converge to the classical result at late times, as quantum 
		effects become negligible with increasing volume.
	\end{itemize}
	
	\subsubsection{Shear Evolution and Anisotropy Suppression}
	
	\begin{figure}[t]
		\centering
		\includegraphics[width=0.45\textwidth]{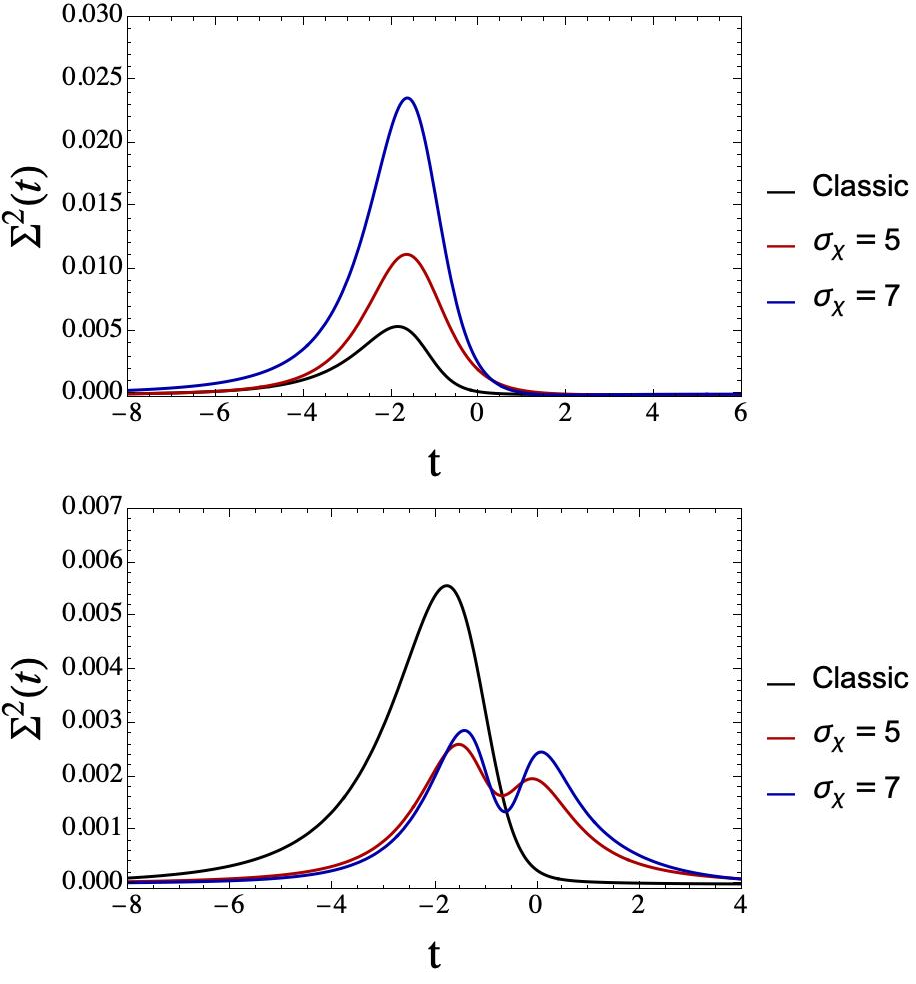}
		\caption{Effective shear $\Sigma^2(t)$ for $\omega = -5$ with cross-correlations (bottom) and without cross-correlations (top). Quantum effects suppress anisotropy: the shear peak is significantly reduced in comparison with the classical one.}
		\label{fig:effective_shear_generic_3d}
	\end{figure}
	
	Figure~\ref{fig:effective_shear_generic_3d} shows the shear $\Sigma^2(t)$ evolution. In contrast to the classical case where shear peaks sharply at the bounce, the effective evolution with cross-correlations shows an anisotropy suppression.
	%: the peak shear decreases with increasing $\sigma_\chi$.
	
	We can also notice post-bounce oscillations in the shear, consistent with Hubble parameter oscillations.
	
	This anisotropy suppression is physically significant: quantum gravity effects near the Planck scale may preferentially smooth anisotropies, potentially explaining why the observable universe is nearly isotropic despite BKL expectations of chaotic anisotropic approach to singularities~\cite{lifshitz1963investigations}.
	
	\subsubsection{Evolution of Quantum Moments}
	
	To understand the microscopic origin of quantum backreaction, we examine the evolution of key quantum moments themselves.
	
\begin{figure}[t]
		\centering
		\includegraphics[width=0.45\textwidth]{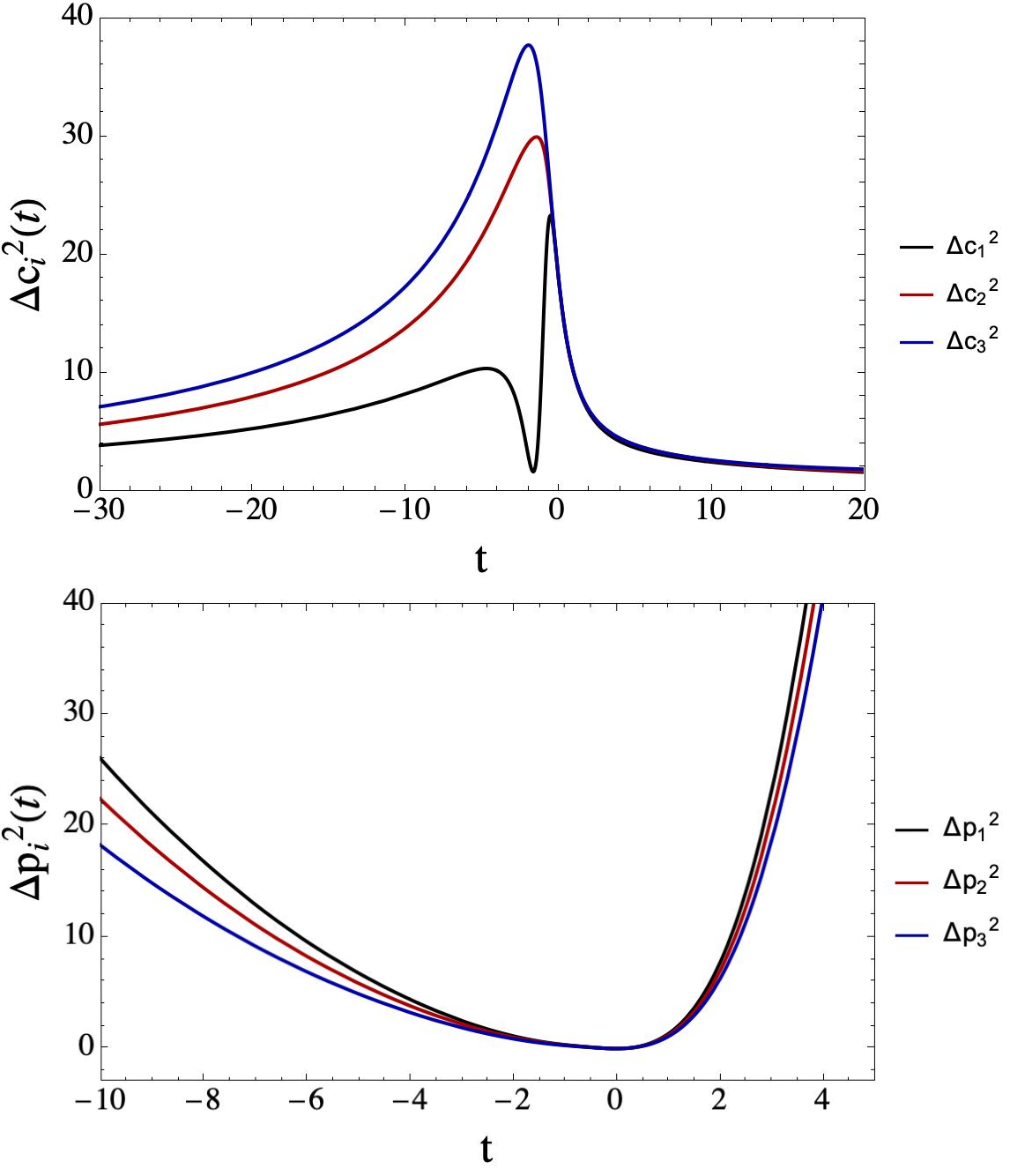}
		\caption{Evolution of quantum dispersions for $\omega = -5$, $\sigma_\chi = 6$. Top: Momentum dispersions $\Delta(c_i^2)$ grow near the bounce as quantum spreading becomes significant, then decay at late times. Bottom: Position dispersions $\Delta(p_i^2)$ decrease near bounce (squeezing) then grow again. The anti-correlation between position and momentum dispersions reflects Heisenberg uncertainty: as one squeezes, the other expands.}
		\label{fig:moments_evolution_generic}
	\end{figure}

	Figure~\ref{fig:moments_evolution_generic} shows the evolution of position and momentum dispersions $\Delta(p_1^2)$ and $\Delta(c_1^2)$ for $\sigma_\chi = 6$.
	
	We can observe that near the bounce, $\Delta(p_i^2)$ decreases (momentum squeezing) while $\Delta(c_i^2)$ increases (position spreading). This reflects quantum state deformation due to strong gravitational effects.
	
\begin{figure}[t]
		\centering
		\includegraphics[width=0.45\textwidth]{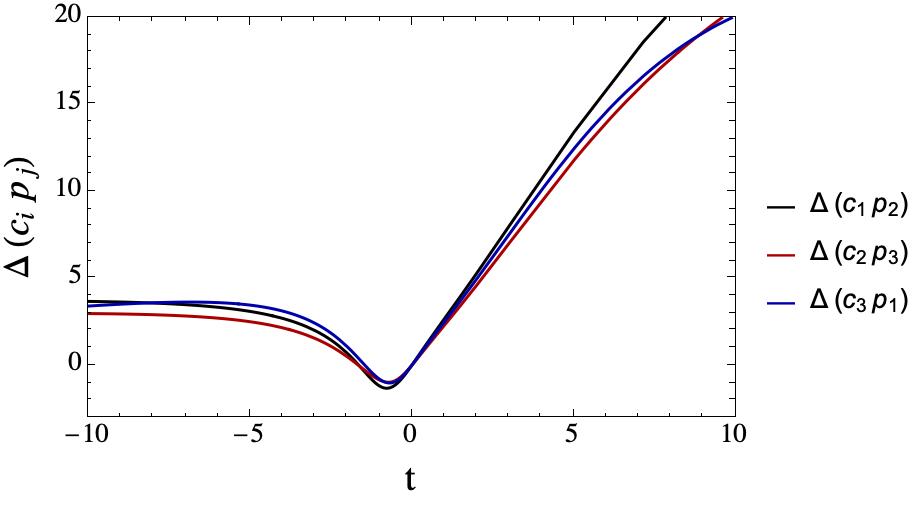}
		\caption{Evolution of representative cross-correlations for $\omega = -5$, $\sigma_\chi = 6$. $\Delta(c_1 p_2)$ (gravitational cross-correlation) grows from zero, having a complex structure near bounce. $\Delta(c_2 p_3)$ and $\Delta(p_1 c_3)$ show similar behavior. All correlations were initialized to zero but develop dynamically due to coupling in the Hamiltonian.}
		\label{fig:correlations_evolution_generic}
	\end{figure}

	Figure~\ref{fig:correlations_evolution_generic} shows evolution of selected cross-correlations: $\Delta(c_1 p_2)$, $\Delta(c_2 p_3)$, and $\Delta(p_1 c_3)$.
	
	It can be observed that all cross-correlations grow from initially zero values, demonstrating they are dynamically generated by Hamiltonian evolution, and that different correlations have different amplitudes, indicating some couple more strongly than others. This analysis provides microscopic confirmation that cross-correlations are physically significant (their magnitudes are comparable to diagonal moments).
	\subsubsection{Quantifying Quantum Corrections}
	
	To quantify the magnitude of quantum backreaction, we define the relative quantum correction to scale factors:
	\begin{equation}
		\delta_{\text{quantum}}(t) = \frac{|a^{\text{eff}}(t) - a^{\text{cl}}(t)|}{a^{\text{cl}}(t)}.
		\label{eq:quantum_correction}
	\end{equation}
	
\begin{figure}[t]
		\centering
		\includegraphics[width=0.4\textwidth]{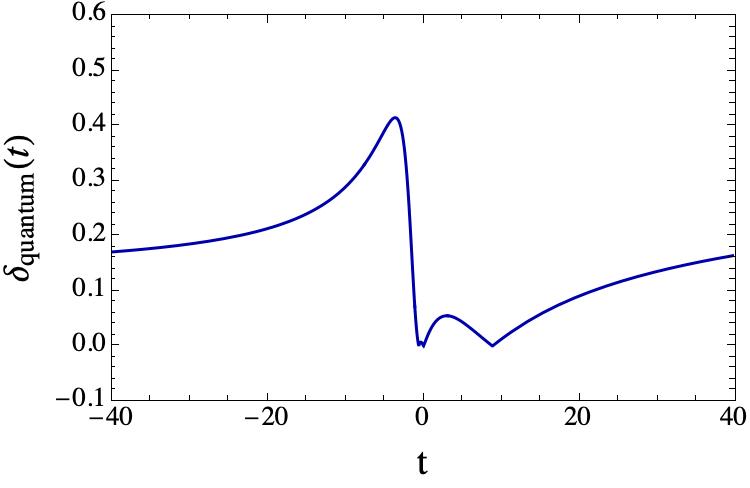}
		\caption{Relative quantum correction $\delta_{\text{quantum}}(t)$ to scale factor for $\omega = -5$. Corrections peak near the bounce, reaching $\delta_{\max} \approx 0.42$ for $\sigma_\chi = 6$.}
		\label{fig:quantum_correction_magnitude}
	\end{figure}

	Figure~\ref{fig:quantum_correction_magnitude} shows $\delta_{\text{quantum}}(t)$ for $\sigma_\chi=6$.
	
	We can notice that the maximum corrections occur at the bounce: $\delta_{\max} \approx 0.42$.
	This confirms that quantum effects are strong enough to be observable and modify dynamics, but weak enough that the semiclassical approximation remains valid.
	
	\subsection{Effective Evolution: Case $\omega = -3/2$}
	\label{subsec:numerical_conformal}
	
	We now examine the conformally invariant case, finding qualitatively similar results with some important distinctions. As in the $\omega = -5$ case we obtain the evolution including cross-correlations. 
	
\begin{figure}[t]
		\centering
		\includegraphics[width=0.45\textwidth]{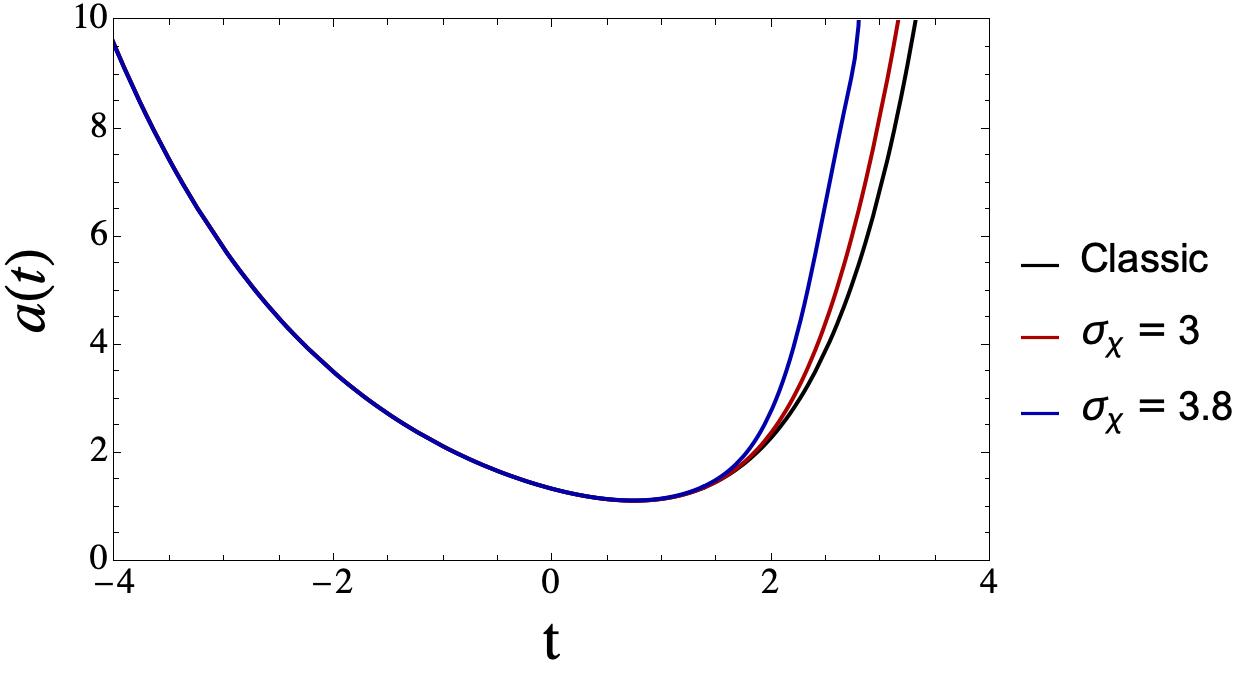}
		\caption{Effective evolution of $a(t)$ for $\omega = -3/2$, $\lambda = 1$. Quantum effects cause scale factors to enter exponential expansion (de Sitter) phase earlier than classically.}
		\label{fig:effective_scales_conformal_3d}
	\end{figure}

	Figure~\ref{fig:effective_scales_conformal_3d} shows the mean scale factor evolution for $\omega = -3/2$. The key qualitative difference from the $\omega = -5$ case is the asymptotic behavior. Quantum effects cause the system to reach de Sitter expansion more rapidly. 
	This acceleration of de Sitter approach is physically interesting for inflationary scenarios: if the universe begins in a quantum state near the bounce with large $\sigma_\chi$, it could enter exponential expansion more rapidly than classical evolution would predict.
	
\begin{figure}[t]
		\centering
		\includegraphics[width=0.45\textwidth]{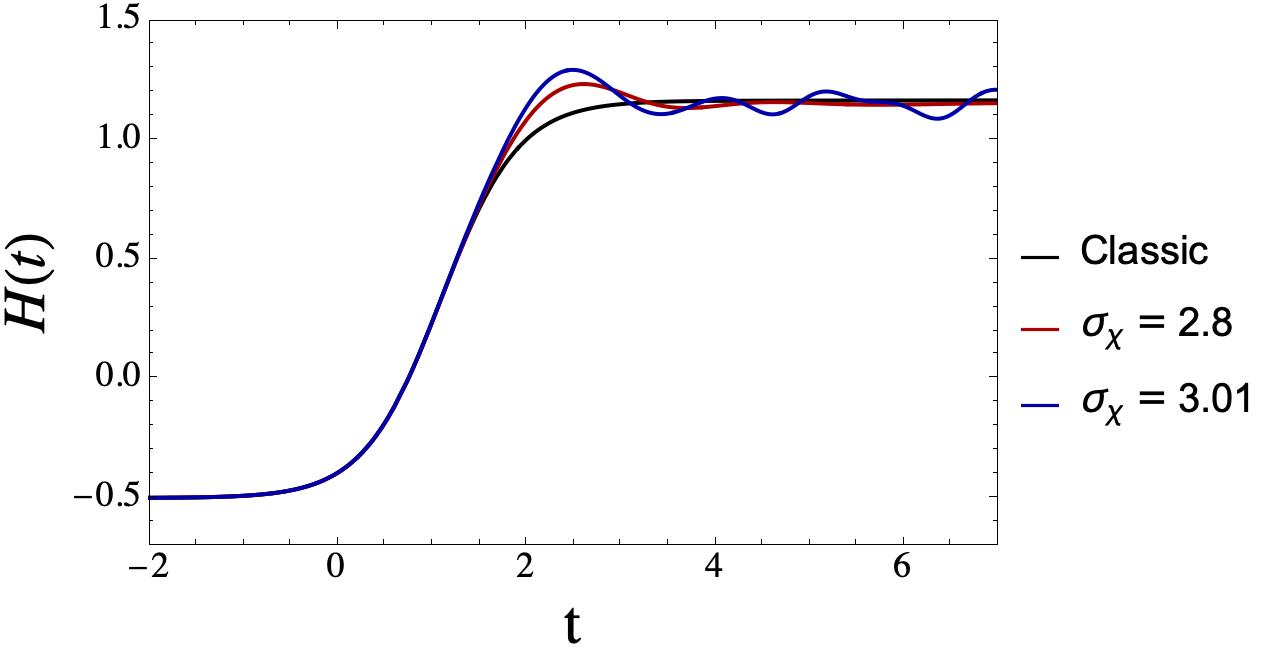}
		\caption{Effective $H(t)$ for $\omega = -3/2, \lambda = 1$, as a function of $\sigma_\chi$. Oscillations appear for $\sigma_\chi \geq 2$. The asymptotic values $H_\infty$ are reached sooner for larger $\sigma_\chi$.}
		\label{fig:effective_hubble_conformal_3d}
	\end{figure}

	Figure~\ref{fig:effective_hubble_conformal_3d} shows the mean Hubble parameter evolution. Post-bounce oscillations appear here as well, with similar characteristics to the $\omega = -5$ case. The similarities suggest oscillations are a generic feature of effective quantum Bianchi I dynamics, not specific to the coupling constant value.
	
	\subsubsection{Energy Density and Shear}
	
\begin{figure}[t]
		\centering
		\includegraphics[width=0.45\textwidth]{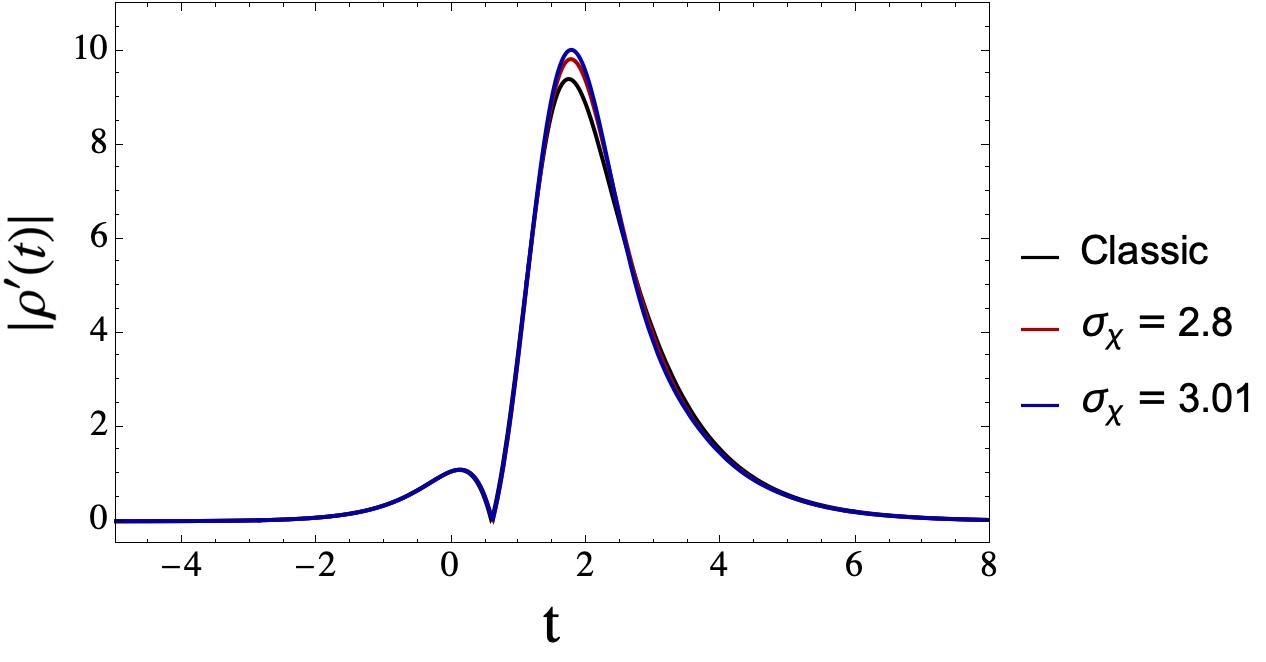}
		\caption{Effective energy density for $\omega = -3/2, \lambda = 1$, various $\sigma_\chi$. The peak enhancement is similar to the $\omega = -5$ case (Fig.~\ref{fig:effective_rho_generic_3d}), confirming this is a generic feature, not specific to the coupling constant value.}
		\label{fig:effective_rho_conformal_3d}
	\end{figure}

	The energy density (Fig.~\ref{fig:effective_rho_conformal_3d}) exhibits a slight increase in the peak maximum, similar to the $\omega = -5$ case, confirming this is a generic quantum effect.
	
\begin{figure}[t]
		\centering
		\includegraphics[width=0.45\textwidth]{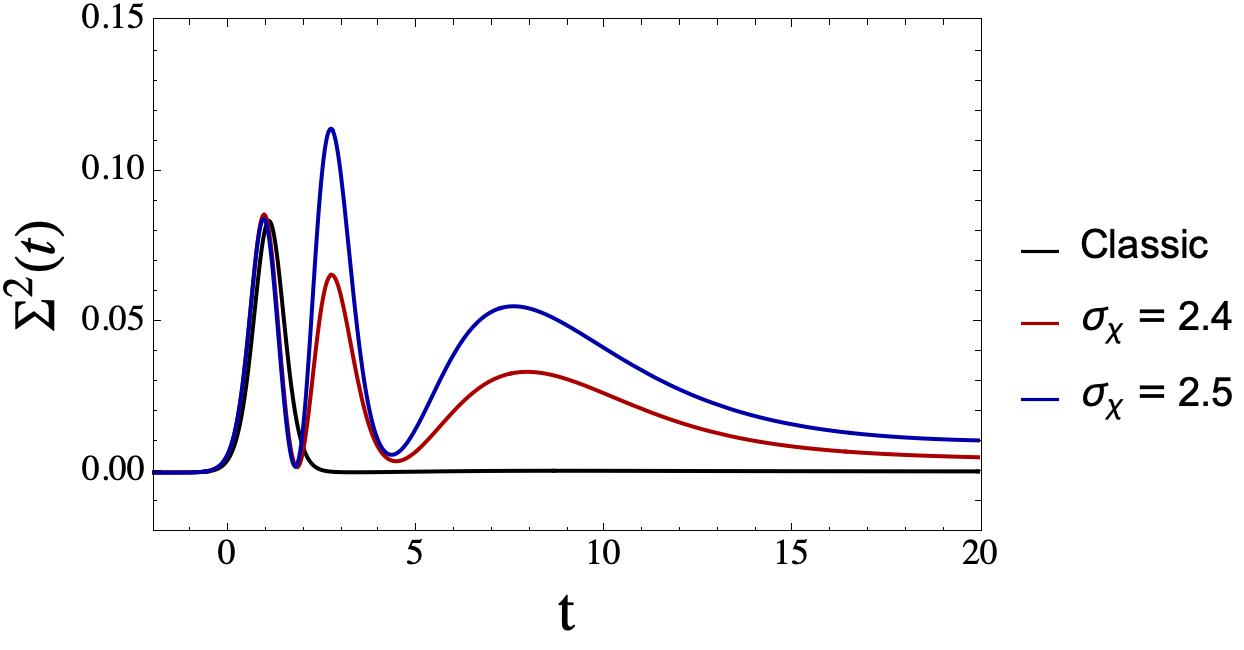}
		\caption{Effective shear for $\omega = -3/2, \lambda = 1$, as a function of $\sigma_\chi$. Unlike the $\omega = -5$ case, here peak shear increases with $\sigma_\chi$, indicating that for the conformally invariant case, quantum effects enhance rather than suppress anisotropy. The physical origin may be related to conformal coupling structure.}
		\label{fig:effective_shear_conformal_3d}
	\end{figure}

	Shear evolution (Fig.~\ref{fig:effective_shear_conformal_3d}) shows a key difference from $\omega = -5$: here, the anisotropy increases with $\sigma_\chi$ rather than decreasing.
	
	This anisotropy enhancement is puzzling and warrants further investigation. We explain this because the conformal constraint~\eqref{eq:conformal_constraint} couples momenta differently, affecting anisotropy generation, the scalar field $\phi$ decays exponentially ($\phi \sim e^{-\lambda t}$), potentially affecting geometric degrees of freedom asymmetrically; and higher-order quantum corrections (beyond our truncation) may be more important for this observable. We plan to analyze this in future work. 
	
	\subsection{Numerical Methods and Error Analysis}
	\label{subsec:numerical_methods}
	
	\subsubsection{Constraint Violation Monitoring}
	
	The Hamiltonian constraint must be satisfied throughout evolution: $\mathcal{H}_{\text{eff}} \approx 0$. For this we analyze
	\begin{equation}
		\nu(t) = \frac{|\mathcal{H}_{\text{eff}}(t)|}{\max_{t'} |\mathcal{H}_{\text{eff}}(t')|}.
		\label{eq:constraint_violation}
	\end{equation}
	This relative normalization is chosen because the constraint is imposed as an initial condition $\mathcal{H}_\text{eff}(0) \approx 0$ and then monitored for drift; dividing by the maximum value provides a scale-independent measure of numerical error growth. In this work the quantity $\nu(t) \leq 1$ is satisfied, which indicates an excellent preservation of the constraint and supports the numerical stability of the integration scheme.
	
	\subsubsection{Heisenberg Uncertainty Verification}
	
	For each degree of freedom, all quantum moments must satisfy the generalized Heisenberg uncertainty relation
	\begin{equation}
		\Xi_i(t) \equiv \Delta(c_i^2) \Delta(p_i^2) - \Delta(c_i p_i)^2 - \frac{\hbar^2}{4} \geq 0.
		\label{eq:uncertainty_check}
	\end{equation}
	
	This is satisfied for both $\omega = -5$ and $\omega = -3/2$ models throughout the evolution, confirming the physical consistency of our quantum states.
	
	\subsubsection{Semiclassicality Monitoring}
	
\begin{figure}[t]
		\centering
		\includegraphics[width=0.45\textwidth]{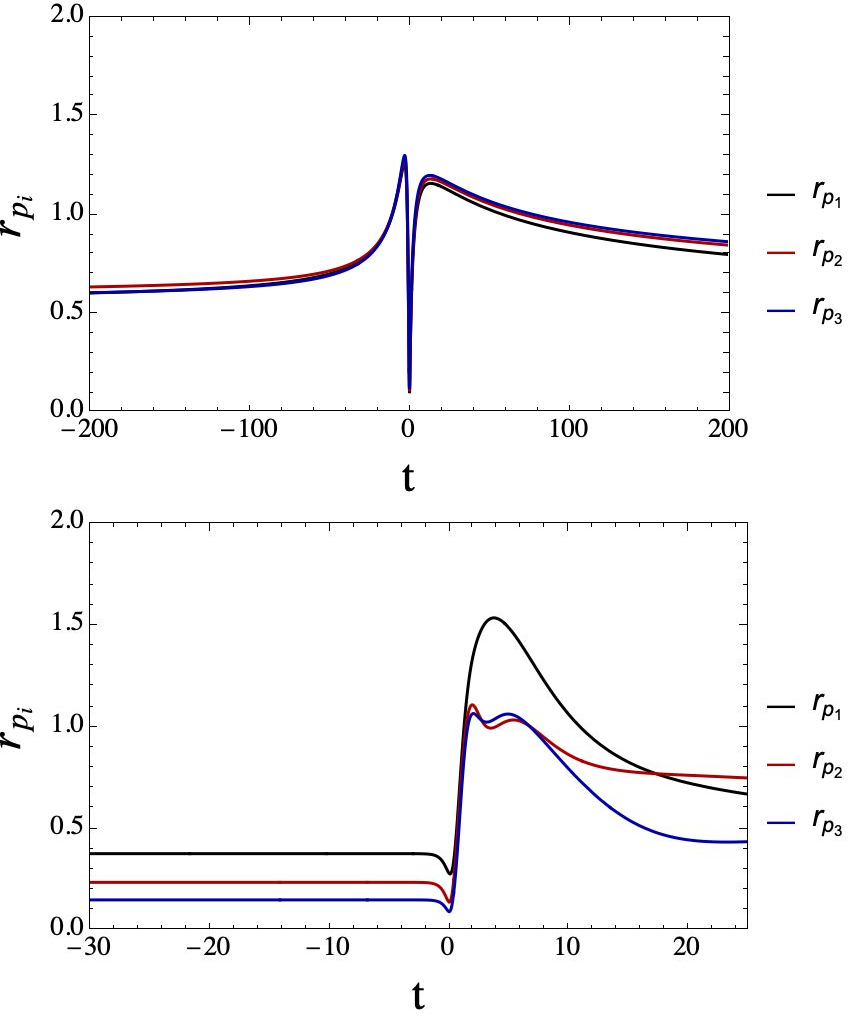}
		\caption{Semiclassicality ratios $r_{p_i}(t)$ for $\omega= -5$, $\sigma_\chi = 8$ (top) and $\omega =-3/2$, $\sigma_\chi = 2.5$ (bottom). All ratios remain small throughout evolution, with maximum values occurring near the bounce. This confirms that quantum fluctuations remain small compared to expectation values, validating the second-order truncation of the effective Hamiltonian.}
		\label{fig:semiclassicality}
	\end{figure}

	We compute the semiclassicality ratios $r_{p_{i}}$ in ~\eqref{eq:semiclassicality_ratios} throughout evolution. Figure~\ref{fig:semiclassicality} shows these ratios for $\omega = -5$, $\sigma_\chi = 8$, and $\omega = -3/2$, $\sigma_\chi = 2.5$ remain small for relatively large quantum states width, justifying the second-order truncation. Near the bounce, ratios increase as quantum effects become more significant, but remain in the regime where semiclassical approximations are valid.
	
	\subsection{Summary of Numerical Results}
	\label{subsec:numerical_summary}
	
	Our numerical investigation establishes several key findings:
	\begin{enumerate}
		\item Cross-correlations are essential: Neglecting quantum correlations between different degrees of freedom produces nonphysical pathologies. Including all 28 cross-correlation terms yields smooth, consistent dynamics. 
		\item Quantum smoothing: For both $\omega = -5$ and $\omega = -3/2$, quantum effects smooth the bounce, spreading it over larger time intervals and preventing collapse to arbitrarily small scales.
		
		\item Post-bounce oscillations: Damped oscillations appear in Hubble parameters and other observables shortly after the bounce. These are quantum remnant effects encoding correlation information.
		
		\item Energy density enhancement: quantum effects cause the energy density to concentrate in a region close to the cosmic bounce similar to the classic evolution, increasing the classical peak.
		
		\item Anisotropy modification: For $\omega = -5$, quantum effects suppress anisotropy. $\omega = -3/2$ they enhance it.
		
		\item De Sitter acceleration ($\omega = -3/2$): Quantum effects cause the system to reach De Sitter expansion phases more rapidly, potentially relevant for inflationary dynamics.
		
		\item Semiclassical validity: For small values of $\sigma_\chi$, semiclassicality ratios remain $< 1.5$, and Heisenberg relations are satisfied, justifying the second-order truncation.
	\end{enumerate}

	\section{Discussion, Conclusions and Outlook}
	\label{sec:discussion}
	
	\subsection{Summary of Main Results}
	We have investigated the effective quantum evolution of 
	Bianchi type~I cosmological models within Brans-Dicke 
	theory, employing a systematic effective Hamiltonian 
	approach that includes expectation values, quantum 
	dispersions, and (crucially) cross-correlation terms 
	between different degrees of freedom. Our principal 
	findings are:
	
	\begin{enumerate}
		
		\item Cross-correlations are essential: Our most 
		important result is the demonstration that quantum 
		cross-correlation terms, coupling different spatial 
		directions and geometry to the scalar field, are 
		absolutely necessary for physically consistent effective 
		dynamics. Neglecting these 28 cross-correlation terms (for our model) 
		produces spurious pathologies and unphysical behavior, 
		while including them yields a smooth and well behaved 
		evolution. This establishes cross-correlations as 
		crucial quantum information carriers, not negligible 
		higher-order corrections. Mathematically, neglecting 
		correlations implicitly assumes the quantum state 
		remains factorized, $|\Psi\rangle = |\psi_1\rangle 
		\otimes |\psi_2\rangle \otimes \cdots$, which is 
		generically violated in interacting systems: the 
		pathologies signal a violation of the Poisson algebra 
		of quantum moments, leading to inconsistencies in the 
		Heisenberg uncertainty relations.
		
		\item Quantum smoothing of bounces: For both 
		$\omega = -5$ and $\omega = -3/2$, quantum backreaction 
		smooths classical bounces. The bounce becomes more 
		gradual, the minimum scale factor increases with 
		$\sigma_\chi$, and the energy density peak is slightly 
		suppressed relative to the classical value, with the 
		suppression growing with $\sigma_\chi$. The unphysical 
		spike that appears when cross-correlations are neglected 
		is entirely removed. We emphasize that for $\omega < 
		-3/2$, the classical BD theory already avoids 
		singularities via the repulsive $-\zeta T^2$ term in 
		Eq.~\eqref{eq:hamiltonian_generic}; quantum effects therefore provide 
		corrections to this nonsingular background.
		
		\item Post-bounce oscillatory remnants: Damped 
		oscillations appear in Hubble parameters and other 
		observables shortly after the bounce. These are direct 
		manifestations of cross-correlation dynamics --- quantum 
		remnant effects that classical evolution cannot produce. 
		They represent transient quantum memory of correlations 
		developed during the Planck-scale bounce phase, and 
		arise because cross-correlations such as $\Delta(c_ip_j)$ 
		and $\Delta(p_ip_j)$ develop complex time dependence 
		near the bounce and back-react on expectation values 
		oscillatorily as the universe expands. Their appearance 
		in both $\omega = -5$ and $\omega = -3/2$ cases suggests 
		they are a generic feature of effective anisotropic 
		quantum cosmology.
		
		\item Anisotropy modification: Quantum effects 
		suppress shear anisotropy for $\omega = -5$, 
		potentially contributing to the observed near-isotropy 
		of the universe despite BKL expectations~\cite{lifshitz1963investigations,belinskii1970oscillatory}. 
		For $\omega = -3/2$, quantum effects instead enhance 
		anisotropy, a puzzling result whose origin may lie in 
		the conformal constraint structure~\eqref{eq:conformal_constraint}, the 
		exponential decay of $\phi$, or higher-order truncation 
		effects. Distinguishing these possibilities requires 
		higher order corrections and a full LQC comparison, 
		which we leave for future work.
		
		\item Accelerated de Sitter approach 
		$(\omega = -3/2)$: For the conformally invariant 
		case, quantum corrections cause the system to reach 
		asymptotic de Sitter expansion more rapidly than 
		classically. This quantum acceleration mechanism may 
		relax fine-tuning requirements on initial conditions 
		for inflation, and connects to the equivalence between 
		$\omega = -3/2$ BD and Palatini $f(R)$ theories~\cite{sotiriou2010f}, 
		suggesting our quantum results may extend to that 
		framework after appropriate variable transformations.
		
		\item Semiclassical validity confirmed: 
		Throughout our parameter range, semiclassicality ratios 
		remain $r < 1.5$ and Heisenberg uncertainty relations 
		are satisfied, validating the second-order truncation. 
		We note that the Gaussian truncation argument is weakest 
		near the bounce where $\mathcal{H}_\mathrm{eff} \to 0$ 
		on-shell, and higher-order corrections $\mathcal{O}
		(\hbar^{3/2})$ may be non-negligible there; we monitor 
		the ratios~\eqref{eq:semiclassicality_ratios} throughout to verify 
		reliability.
		
	\end{enumerate}
	
	\subsection{Comparison with Previous Work}
	
	Our results complement and extend previous work on 
	quantum cosmology along three axes.
	
	Loop quantum cosmology. Loop quantum Bianchi~I 
	models in pure GR~\cite{ashtekar2009loop, chiou2007loop} 
	and in Brans-Dicke theory~\cite{sharma2025quantum, zhang2012loop} 
	demonstrate singularity resolution with bounded 
	curvature invariants, bounces at $p_i \sim \ell_\mathrm{Pl}^2$, 
	and preservation of anisotropic structure. Our effective 
	approach reproduces these qualitative features --- finite 
	energy density, quantum-determined bounce scale set by 
	$\sigma_\chi$, anisotropy preservation, and smooth 
	transition through the bounce --- using a complementary 
	method that provides fully dynamical equations for all 
	quantum moments. The crucial difference is that 
	reference~\cite{sharma2025quantum} uses coherent state peaking to 
	reduce to modified classical equations, whereas we 
	evolve the complete 48-dimensional extended phase space. 
	A rough estimate from our numerics with $\sigma_\chi 
	\sim 5$--$10$ yields bounce densities $\rho_\mathrm{max} 
	\sim 0.3$--$0.5$ in natural units, comparable to LQC 
	predictions~\cite{ashtekar2009loop}, though a 
	precise comparison requires matched initial conditions 
	which we leave for future work.
	
	Effective approaches for isotropic models. 
	Reference~\cite{bojowald2007effective} developed effective 
	equations for FLRW models with matter, finding bounce 
	densities $\rho_\mathrm{max} \approx 0.41\,\rho_\mathrm{Pl}$ 
	and quantum corrections decreasing with volume. Our 
	Bianchi~I results extend these findings to anisotropic 
	models with modified gravity. The crucial qualitative 
	difference is that cross-correlations are negligible in 
	FLRW by symmetry, but dynamically essential in Bianchi~I. 
	This aligns with recent Mixmaster (Bianchi~IX) effective 
	results~\cite{hernandez2024singularity}, where 
	cross-correlations similarly proved necessary for 
	physical consistency, suggesting that their importance 
	is a universal feature of anisotropic quantum cosmology 
	rather than a model-specific artifact.
	
	Observational implications. 
	Cross-correlations between gravitational and matter 
	degrees of freedom may have broader phenomenological 
	consequences. Near Planck-scale energies, correlations 
	$\Delta(h_{ij}\phi)$ between metric perturbations and 
	the inflaton could modify primordial power spectra and 
	non-Gaussianity, requiring extension of our homogeneous 
	analysis to perturbations. Post-bounce oscillations, 
	if present in the early universe, could leave imprints 
	in primordial gravitational waves or CMB anomalies. 
	The BD theory with $\omega < -3/2$ has negative kinetic 
	energy ($\zeta < 0$), analogous to phantom scalar 
	fields~\cite{caldwell2002phantom}, for which our effective approach 
	provides a consistent semiclassical treatment by working 
	with real expectation values rather than operators; 
	however, validity in strongly phantom-like regimes 
	(very negative $\omega$) requires further investigation.
	
	\subsection{Future Directions}
	
	Several natural extensions follow from this work:
	
	\begin{itemize}
		
		\item Higher-order corrections: Computing 
		$\mathcal{O}(\hbar^{3/2})$ terms in the effective 
		Hamiltonian to quantify truncation errors near the 
		bounce and extend the validity range, particularly 
		for the shear anisotropy in the $\omega = -3/2$ case.
		
		\item Non-Gaussian and squeezed states: 
		Exploring coherent state superpositions and Wigner 
		function evolution to assess sensitivity to the 
		initial quantum state beyond factorized Gaussians.
		
		\item Other Bianchi models: Applying our 
		cross-correlation-inclusive approach to Bianchi~II, 
		VIII, and IX models to determine whether the 
		universality of cross-correlation importance extends 
		across Bianchi types.
		
		\item Detailed LQC comparison: Once loop 
		quantization of BD Bianchi~I is further 
		developed~\cite{sharma2025quantum}, performing a detailed 
		quantitative comparison with matched initial 
		conditions to identify regimes of agreement and 
		disagreement.
		
		\item Cosmological perturbations and 
		observational signatures: Extending the analysis 
		to perturbations around the Bianchi~I background 
		to compute primordial power spectra and 
		non-Gaussianity signatures from cross-correlations, 
		with a view toward comparison with Planck, LIGO, 
		and future experiments.
		
	\end{itemize}
	
	\subsection{Concluding Remarks}
	
	This work demonstrates that effective quantum methods, 
	when properly accounting for all relevant quantum 
	degrees of freedom and especially cross-correlations, 
	provide a powerful tool for studying quantum cosmology 
	in scenarios where full quantization is intractable. 
	The post-bounce oscillations we have identified 
	represent a new type of quantum remnant effect: 
	transient but observable consequences of Planck-scale 
	quantum structure that persist into semiclassical 
	regimes and may have left imprints detectable today. 
	The interplay between classical modified-gravity 
	effects --- BD theory with $\omega < -3/2$ already 
	avoiding singularities --- and quantum corrections 
	that smooth, oscillate, and modify anisotropy 
	illustrates the rich phenomenology possible when 
	these approaches are combined. Disentangling 
	classical from quantum contributions will be 
	crucial for any future observational tests of 
	quantum gravity.
	%%%%%%%%%%%%%%%%%%%%%%%%%%%%%%%%%%%%%%%%%%
	\begin{acknowledgments}
		G.S.H. acknowledges the financial support provided by SECIHTI through a doctoral scholarship. H.H.H. acknowledges SECIHTI Sabbatical Grant 2025. This work was supported by CONAHCYT/SECIHTI Grant CBF-2023-2024-1937. We thank J. Arroyo for helpful discussions. 
	\end{acknowledgments}
	
	\section*{Data Availability}
	The data supporting the findings in this article were generated using Mathematica and are available at \cite{data_repo}.
	
	\appendix
	
	\section{ADM Formulation and Ashtekar Variables}
	\label{app:adm}
	
	In this appendix, we provide technical details of the ADM decomposition and derivation of the Hamiltonian formulation in Ashtekar variables for Brans-Dicke theory, complementing the summary in Sec.~\ref{subsec:hamiltonian}.
	
	\subsection{ADM Decomposition of Spacetime}
	
	The Arnowitt-Deser-Misner (ADM) formulation~\cite{arnowitt2008dynamics} provides a Hamiltonian description of general relativity by decomposing spacetime into space and time.
	
	\subsubsection{Foliation and Line Element}
	
	Consider a spacetime manifold $\mathcal{M}$ foliated by spacelike hypersurfaces $\Sigma_t$ labeled by time coordinate $t$. The line element takes the form
	\begin{equation}
		ds^2 = -N^2 dt^2 + q_{ab}(dx^a + N^a dt)(dx^b + N^b dt),
		\label{eq:adm_metric}
	\end{equation}
	where:
	\begin{itemize}
		\item $N(t, x^a)$ is the lapse function, measuring proper time between adjacent hypersurfaces
		\item $N^a(t, x^b)$ is the shift vector, relating spatial coordinates on adjacent hypersurfaces
		\item $q_{ab}(t, x^c)$ is the spatial metric on $\Sigma_t$
	\end{itemize}
	
	The inverse spacetime metric is
	\begin{equation}
		g^{\mu\nu} = \begin{pmatrix}
			-N^{-2} & N^{-2} N^a \\
			N^{-2} N^b & q^{ab} - N^{-2} N^a N^b
		\end{pmatrix}.
	\end{equation}
	
	\subsubsection{Extrinsic Curvature}
	
	The extrinsic curvature $K_{ab}$ of $\Sigma_t$ embedded in $\mathcal{M}$ is defined as
	\begin{equation}
		K_{ab} = \frac{1}{2N}\left( \dot{q}_{ab} - \mathcal{L}_{\vec{N}} q_{ab} \right) = \frac{1}{2N}(\dot{q}_{ab} - D_a N_b - D_b N_a),
		\label{eq:extrinsic_curvature}
	\end{equation}
	where $D_a$ is the covariant derivative compatible with $q_{ab}$, and $\mathcal{L}_{\vec{N}}$ denotes Lie derivative along the shift.
	
	The trace is $K = q^{ab} K_{ab}$.
	
	\subsubsection{ADM Action for General Relativity}
	
	Substituting the ADM decomposition into the Einstein-Hilbert action
	\begin{equation}
		S_{EH} = \frac{1}{16\pi G} \int d^4x \sqrt{-g} \, R,
	\end{equation}
	and integrating by parts yields (setting $16\pi G = 1$)
	\begin{equation}
		S_{ADM} = \int dt \int_{\Sigma} d^3x \, N\sqrt{q} \left[ {}^{(3)}R + K_{ab}K^{ab} - K^2 \right],
		\label{eq:adm_action_gr}
	\end{equation}
	where ${}^{(3)}R$ is the Ricci scalar of the spatial metric $q_{ab}$.
	
	\subsubsection{Canonical Variables and Hamiltonian}
	
	Define the canonical momentum conjugate to $q_{ab}$:
	\begin{equation}
		\pi^{ab} = \frac{\delta L}{\delta \dot{q}_{ab}} = \sqrt{q}(K^{ab} - K q^{ab}).
	\end{equation}
	
	The Hamiltonian is
	\begin{equation}
		H_{ADM} = \int_{\Sigma} d^3x (N \mathcal{C} + N^a \mathcal{C}_a),
		\label{eq:adm_hamiltonian}
	\end{equation}
	where the Hamiltonian and diffeomorphism constraints are
	\begin{subequations}
		\begin{align}
			\mathcal{C} &= G_{abcd} \pi^{ab} \pi^{cd} - \sqrt{q} \, {}^{(3)}R \approx 0, \label{eq:hamiltonian_constraint_adm} \\
			\mathcal{C}_a &= -2 D_b \pi^b_a \approx 0, \label{eq:diffeomorphism_constraint_adm}
		\end{align}
	\end{subequations}
	with the DeWitt supermetric
	\begin{equation}
		G_{abcd} = \frac{1}{2\sqrt{q}}(q_{ac}q_{bd} + q_{ad}q_{bc} - 2q_{ab}q_{cd}).
	\end{equation}
	
	\subsection{Ashtekar-Barbero Variables}
	
	The Ashtekar formulation~\cite{ashtekar1987new} recasts the gravitational phase space in terms of an $SU(2)$ connection and its conjugate.
	
	\subsubsection{Densitized Triad and Connection}
	
	Introduce a spatial triad $e^i_a(x)$ such that
	\begin{equation}
		q_{ab} = \delta_{ij} e^i_a e^j_b,
	\end{equation}
	with inverse (co-triad) $e^a_i$ satisfying $e^a_i e^j_a = \delta^j_i$ and $e^a_i e^b_i = \delta^b_a$.
	
	Define the densitized triad:
	\begin{equation}
		E^a_i = \sqrt{q} \, e^a_i, \quad \text{with} \quad \det(E^a_i) = \sqrt{q}.
	\end{equation}
	
	The Ashtekar-Barbero connection is
	\begin{equation}
		A^i_a = \Gamma^i_a + \gamma K^i_a,
		\label{eq:ashtekar_connection}
	\end{equation}
	where:
	\begin{itemize}
		\item $\Gamma^i_a$ is the spin connection compatible with $e^i_a$:
		\begin{equation}
			\Gamma^i_a = -\epsilon^{ijk} e_j^b (\partial_a e_{bk} - \partial_b e_{ak}),
		\end{equation}
		
		\item $K^i_a = K_{ab} e^{bi}$ is the extrinsic curvature in triad components
		
		\item $\gamma$ is the Barbero-Immirzi parameter~\cite{barbero1994real,immirzi1997real}
	\end{itemize}
	
	\subsubsection{Canonical Structure}
	
	The canonical Poisson brackets are
	\begin{equation}
		\{A^i_a(x), E^b_j(y)\} = 8\pi G \gamma \delta^i_j \delta^b_a \delta^{(3)}(x,y).
		\label{eq:poisson_ashtekar}
	\end{equation}
	
	\subsubsection{Hamiltonian in Ashtekar Variables}
	
	The gravitational Hamiltonian becomes~\cite{thiemann2008modern}
	\begin{equation}
		H_g = \int d^3x (N \mathcal{C}_g + N^a \mathcal{C}_{g,a} + \Lambda^i \mathcal{G}_i),
	\end{equation}
	with constraints
	\begin{subequations}
		\begin{align}
			\mathcal{C}_g &= -\frac{1}{8\pi G\gamma^2} \frac{\epsilon^{ijk} E^a_i E^b_j}{\sqrt{\det E}} \left[ F^k_{ab} - (1 + \gamma^2)\epsilon^k_{mn} K^m_a K^n_b \right] \approx 0, \\
			\mathcal{C}_{g,a} &= F^i_{ab} E^b_i \approx 0, \\
			\mathcal{G}_i &= D^{(A)}_a E^a_i = \partial_a E^a_i + \epsilon_{ijk} A^j_a E^{ak} \approx 0,
		\end{align}
	\end{subequations}
	where $F^i_{ab} = 2\partial_{[a} A^i_{b]} + \epsilon^i_{jk} A^j_a A^k_b$ is the curvature of $A^i_a$.
	
	\subsection{Brans-Dicke Theory in ADM Variables}
	
	\subsubsection{ADM Decomposition of BD Action}
	
	The Brans-Dicke action~\eqref{eq:bd_action} in ADM form becomes
	\begin{equation}
		\begin{split}
			S_{BD} = \int dt \int_{\Sigma} d^3x \, N\sqrt{q} \Big[
			& \phi \left( {}^{(3)}R + K_{ab}K^{ab} - K^2 \right) \\
			& - \frac{\omega}{\phi N^2} \left( \dot{\phi} - N^a \partial_a \phi \right)^2 \\
			&+ \frac{\omega}{\phi} q^{ab} \partial_a \phi \partial_b \phi \\
			& + \frac{2}{N} \left( \dot{\phi} - N^a \partial_a \phi \right) K \Big].
		\end{split}
	\end{equation}
	
	\subsubsection{Canonical Variables for Scalar Field}
	
	The scalar field has canonical momentum
	\begin{equation}
		p_\phi = \frac{\delta L_{BD}}{\delta \dot{\phi}} = \frac{2\sqrt{q}}{N}\left[ K \phi - \frac{\omega}{\phi}(\dot{\phi} - N^a \partial_a \phi) \right],
	\end{equation}
	with Poisson bracket $\{\phi(x), p_\phi(y)\} = \delta^{(3)}(x,y)$.
	
	\subsubsection{BD Hamiltonian Constraint}
	
	After Legendre transformation and expressing $K_{ab}$ in terms of momenta, the Hamiltonian constraint becomes (for spatially homogeneous configurations)
	\begin{equation}
		\mathcal{H}_{BD} = \mathcal{H}_g[\phi] + \mathcal{H}_\phi[p_\phi, \phi] \approx 0,
	\end{equation}
	where the precise form depends on $\omega$. For $\omega \neq -3/2$, this reduces to Eq.~\eqref{eq:hamiltonian_generic} after symmetry reduction to Bianchi I.
	
	\subsection{Bianchi I Symmetry Reduction}
	
	\subsubsection{Diagonal Metric Ansatz}
	
	For Bianchi I, the spatial metric is diagonal and homogeneous:
	\begin{equation}
		q_{ab} dx^a dx^b = a_1^2(t) dx_1^2 + a_2^2(t) dx_2^2 + a_3^2(t) dx_3^2.
	\end{equation}
	
	Choose a fiducial flat metric $\mathring{q}_{ab}$ with $\mathring{q}_{ab} dx^a dx^b = dx_1^2 + dx_2^2 + dx_3^2$.
	
	\subsubsection{Fiducial Cell and Volume}
	
	Since spacetime is noncompact, integrals over $\Sigma$ diverge. We introduce a fiducial cell $\mathcal{V}_0$ with edges $L_i$ and volume $V_0 = L_1 L_2 L_3$. All integrations are restricted to $\mathcal{V}_0$.
	
	\subsubsection{Reduced Canonical Variables}
	
	For Bianchi I with fiducial triad $\mathring{e}^a_i = \delta^a_i$, the Ashtekar variables take the form given in Eq.~\eqref{eq:ashtekar_vars}:
	\begin{equation}
		A^i_a = c_i (L_i)^{-1} \mathring{e}^i_a, \quad E^a_i = p_i L_i V_0^{-1} \sqrt{\mathring{q}} \, \mathring{e}^a_i,
	\end{equation}
	with $(c_i, p_i)$ time-dependent functions satisfying~\eqref{eq:poisson_cp}.
	
	\subsubsection{Spin Connection for Bianchi I}
	
	For diagonal Bianchi I, the spin connection vanishes: $\Gamma^i_a = 0$. This simplifies the curvature:
	\begin{equation}
		F^i_{ab} = \gamma \left( 2\partial_{[a} K^i_{b]} + \gamma \epsilon^i_{jk} K^j_a K^k_b \right).
	\end{equation}
	
	The vanishing of the spin connection $\Gamma^i_a = 0$ follows from the diagonal and homogeneous form of the Bianchi I metric: since $e^i_a = \delta^i_a$ for the diagonal triad, the structure equation $de^i + \Gamma^i_{\ j}\wedge e^j = 0$ is trivially satisfied with $\Gamma^i_a = 0$. As a consequence, the Ashtekar--Barbero connection~\eqref{eq:ashtekar_connection} reduces to $A^i_a = \gamma K^i_a$, so that the reduced canonical variables $c_i$ defined in Eq.~\eqref{eq:ashtekar_vars} are directly related to the extrinsic curvature via $c_i = \gamma\,a_i H_i$ (no orientation factor). This makes the physical interpretation of $c_i$ particularly transparent.
	
	In terms of $(c_i, p_i)$, after imposing homogeneity and integrating over $\mathcal{V}_0$, the Hamiltonian constraint reduces to Eq.~\eqref{eq:hamiltonian_generic} for $\omega \neq -3/2$ and Eq.~\eqref{eq:hamiltonian_conformal} for $\omega = -3/2$.
	
	\subsubsection{Relation to Scale Factors}
	
	The scale factors are related to $p_i$ by Eq.~\eqref{eq:scale_factors}. These can be derived from
	\begin{equation}
		a_i = \frac{(\det E)^{1/3}}{|E^i_i|^{1/2}} = \frac{(p_1 p_2 p_3)^{1/6}}{|p_i|^{1/2}},
	\end{equation}
	with appropriate sign choices for orientation.
	
	\section{Explicit Equations of Motion}
	\label{app:eom}
	
	We provide the complete equations of motion for the effective quantum systems studied in Secs.~\ref{subsec:numerical_generic} and~\ref{subsec:numerical_conformal}.
	
	\subsection{Effective Poisson Algebra}
	
	The quantum moments satisfy a closed Poisson algebra derived from canonical commutation relations~\cite{bojowald2012quantum}. For a single degree of freedom $(q, p)$:
	\begin{subequations}
		\begin{align}
			\{\Delta(q^2), \Delta(p^2)\} &= 4\Delta(qp), \\
			\{\Delta(q^2), \Delta(qp)\} &= 2\Delta(q^2), \\
			\{\Delta(p^2), \Delta(qp)\} &= -2\Delta(p^2), \\
			\{q, \Delta(q^2)\} &= 2\Delta(qp), \\
			\{p, \Delta(q^2)\} &= 0, \\
			\{q, \Delta(p^2)\} &= 0, \\
			\{p, \Delta(p^2)\} &= -2\Delta(qp), \\
			\{q, \Delta(qp)\} &= \Delta(p^2), \\
			\{p, \Delta(qp)\} &= -\Delta(q^2).
		\end{align}
	\end{subequations}
	
	For multiple degrees of freedom, additional brackets involve cross-correlations:
	\begin{subequations}
		\begin{align}
			\{q_i, \Delta(q_j p_k)\} &= \delta_{ij} \Delta(p_k^2) + \delta_{ik} \Delta(q_j p_k), \\
			\{p_i, \Delta(q_j p_k)\} &= -\delta_{ij} \Delta(q_k^2) - \delta_{ik} \Delta(q_j q_k), \\
			\{\Delta(q_i p_i), \Delta(q_j p_j)\} 
			&= \Delta(p_i p_j)\Delta(q_i q_j) - \Delta(q_i p_j)\Delta(p_i q_j), \nonumber \\
			& \hspace{0.5cm} \text{for} \hspace{0.3cm} i\neq j. 
		\end{align}
	\end{subequations}
	
	\subsection{Equations for Expectation Values: $\omega \neq -3/2$}
	
	The evolution of expectation values is given by $\dot{f} = \{f, H_{\text{eff}}\}$. For our system:
	\begin{subequations}
		\label{eq:eom_expectations_generic_full}
		\begin{align}
			\dot{c}_1 &= \frac{\partial H_{\text{eff}}}{\partial p_1} = \frac{\partial H_{\text{BD}}}{\partial p_1} + \sum_{a,b} \frac{\partial^2 H_{\text{BD}}}{\partial p_1 \partial x_{ab}} \Delta(x_{ab}), \\
			\dot{c}_2 &= \frac{\partial H_{\text{eff}}}{\partial p_2} = \frac{\partial H_{\text{BD}}}{\partial p_2} + \sum_{a,b} \frac{\partial^2 H_{\text{BD}}}{\partial p_2 \partial x_{ab}} \Delta(x_{ab}), \\
			\dot{c}_3 &= \frac{\partial H_{\text{eff}}}{\partial p_3} = \frac{\partial H_{\text{BD}}}{\partial p_3} + \sum_{a,b} \frac{\partial^2 H_{\text{BD}}}{\partial p_3 \partial x_{ab}} \Delta(x_{ab}), \\
			\dot{p}_1 &= -\frac{\partial H_{\text{eff}}}{\partial c_1} = -\frac{\partial H_{\text{BD}}}{\partial c_1} - \sum_{a,b} \frac{\partial^2 H_{\text{BD}}}{\partial c_1 \partial x_{ab}} \Delta(x_{ab}), \\
			\dot{p}_2 &= -\frac{\partial H_{\text{eff}}}{\partial c_2} = -\frac{\partial H_{\text{BD}}}{\partial c_2} - \sum_{a,b} \frac{\partial^2 H_{\text{BD}}}{\partial c_2 \partial x_{ab}} \Delta(x_{ab}), \\
			\dot{p}_3 &= -\frac{\partial H_{\text{eff}}}{\partial c_3} = -\frac{\partial H_{\text{BD}}}{\partial c_3} - \sum_{a,b} \frac{\partial^2 H_{\text{BD}}}{\partial c_3 \partial x_{ab}} \Delta(x_{ab}), \\
			\dot{\phi} &= \frac{\partial H_{\text{eff}}}{\partial p_\phi} = \frac{\partial H_{\text{BD}}}{\partial p_\phi} + \sum_{a,b} \frac{\partial^2 H_{\text{BD}}}{\partial p_\phi \partial x_{ab}} \Delta(x_{ab}), \\
			\dot{p}_\phi &= -\frac{\partial H_{\text{eff}}}{\partial \phi} = -\frac{\partial H_{\text{BD}}}{\partial \phi} - \sum_{a,b} \frac{\partial^2 H_{\text{BD}}}{\partial \phi \partial x_{ab}} \Delta(x_{ab}),
		\end{align}
	\end{subequations}
	where $x_{ab}$ collectively denotes all phase space variables $(c_i, p_i, \phi, p_\phi)$, and the sums run over all quantum moments.
	
	Explicitly, from Eq.~\eqref{eq:eom_generic} plus quantum corrections:
	\begin{equation}
		\begin{split}
			\dot{c}_1 = & \frac{1}{\phi\sqrt{p}} \left[ -c_1(c_2 p_2 + c_3 p_3 - 2\zeta T) + \frac{S - \zeta T^2}{2p_1} \right] \\
			& + \frac{1}{\sqrt{p}} \left[ \frac{p_1^2 \zeta}{\phi} \frac{\partial}{\partial p_1}\Delta(c_1^2) + \frac{2p_1 \zeta}{\sqrt{p}} \frac{\partial}{\partial p_1}\Delta(c_1 p_\phi) + \cdots \right],
		\end{split}
	\end{equation}
	and similarly for other expectation values. The full expressions are lengthy; we implement them numerically using Mathematica.
	
	\subsection{Equations for Quantum Moments: $\omega \neq -3/2$}
	
	For quantum dispersions (diagonal moments), the evolution is:
	\begin{subequations}
		\label{eq:eom_dispersions_generic}
		\begin{align}
			\frac{1}{2}\dot{\Delta}(c_1^2) &= \frac{\partial^2 H_{\text{eff}}}{\partial p_1^2} \Delta(c_1 p_1) + \frac{\partial^2 H_{\text{eff}}}{\partial c_1 \partial p_1} \Delta(c_1^2) \nonumber \\
			&+ \sum_{j \neq 1} \left[ \frac{\partial^2 H_{\text{eff}}}{\partial p_1 \partial c_j} \Delta(c_1 c_j) + \frac{\partial^2 H_{\text{eff}}}{\partial p_1 \partial p_j} \Delta(c_1 p_j) \right] \notag \\
			& \quad + \frac{\partial^2 H_{\text{eff}}}{\partial p_1 \partial \phi} \Delta(c_1 \phi) + \frac{\partial^2 H_{\text{eff}}}{\partial p_1 \partial p_\phi} \Delta(c_1 p_\phi), \\
			\frac{1}{2}\dot{\Delta}(p_1^2) &= -\frac{\partial^2 H_{\text{eff}}}{\partial c_1^2} \Delta(c_1 p_1) - \frac{\partial^2 H_{\text{eff}}}{\partial c_1 \partial p_1} \Delta(p_1^2) \notag \\
			& - \sum_{j \neq 1} \left[ \frac{\partial^2 H_{\text{eff}}}{\partial c_1 \partial c_j} \Delta(p_1 c_j) + \frac{\partial^2 H_{\text{eff}}}{\partial c_1 \partial p_j} \Delta(p_1 p_j) \right] \notag \\
			& \quad - \frac{\partial^2 H_{\text{eff}}}{\partial c_1 \partial \phi} \Delta(p_1 \phi) - \frac{\partial^2 H_{\text{eff}}}{\partial c_1 \partial p_\phi} \Delta(p_1 p_\phi),
		\end{align}
	\end{subequations}
	with analogous equations for $\Delta(c_2^2)$, $\Delta(p_2^2)$, $\Delta(c_3^2)$, $\Delta(p_3^2)$, $\Delta(\phi^2)$, $\Delta(p_\phi^2)$.
	For covariances (e.g., $\Delta(c_1 p_1)$):
	\begin{equation}
		\begin{split}
			\dot{\Delta}(c_1 p_1) = & -\frac{\partial^2 H_{\text{eff}}}{\partial c_1^2} \Delta(c_1^2) + \frac{\partial^2 H_{\text{eff}}}{\partial p_1^2} \Delta(p_1^2) \\
			& - \sum_{j \neq 1} \left[ \frac{\partial^2 H_{\text{eff}}}{\partial c_1 \partial c_j} \Delta(c_1 c_j) + \frac{\partial^2 H_{\text{eff}}}{\partial c_1 \partial p_j} \Delta(c_1 p_j) \right] \\
			& + \sum_{j \neq 1} \left[ \frac{\partial^2 H_{\text{eff}}}{\partial p_1 \partial c_j} \Delta(p_1 c_j) + \frac{\partial^2 H_{\text{eff}}}{\partial p_1 \partial p_j} \Delta(p_1 p_j) \right] \\
			& - \frac{\partial^2 H_{\text{eff}}}{\partial c_1 \partial \phi} \Delta(c_1 \phi) + \frac{\partial^2 H_{\text{eff}}}{\partial p_1 \partial \phi} \Delta(p_1 \phi) \\
			& - \frac{\partial^2 H_{\text{eff}}}{\partial c_1 \partial p_\phi} \Delta(c_1 p_\phi) + \frac{\partial^2 H_{\text{eff}}}{\partial p_1 \partial p_\phi} \Delta(p_1 p_\phi).
		\end{split}
		\label{eq:eom_covariance_example}
	\end{equation}
	
	For cross-correlations (e.g., $\Delta(c_1 p_2)$):
	\begin{equation}
		\begin{split}
			\dot{\Delta}(c_1 p_2) = & \frac{\partial^2 H_{\text{eff}}}{\partial p_1^2} \Delta(p_1 p_2) - \frac{\partial^2 H_{\text{eff}}}{\partial c_2^2} \Delta(c_1 c_2) \\
			& + \frac{\partial^2 H_{\text{eff}}}{\partial c_1 \partial p_1} \Delta(c_1 p_2) - \frac{\partial^2 H_{\text{eff}}}{\partial c_1 \partial c_2} \Delta(c_1^2) \\
			& + \frac{\partial^2 H_{\text{eff}}}{\partial p_1 \partial c_2} [\Delta(c_2 p_2) - \Delta(c_1 p_1)] + \frac{\partial^2 H_{\text{eff}}}{\partial p_1 \partial p_2} \Delta(p_2^2) \\
			& + \sum_{k \neq 1,2} \left[ \frac{\partial^2 H_{\text{eff}}}{\partial p_1 \partial c_k} \Delta(p_2 c_k) + \frac{\partial^2 H_{\text{eff}}}{\partial p_1 \partial p_k} \Delta(p_2 p_k) \right] \\
			& - \sum_{k \neq 1,2} \left[ \frac{\partial^2 H_{\text{eff}}}{\partial c_2 \partial c_k} \Delta(c_1 c_k) + \frac{\partial^2 H_{\text{eff}}}{\partial c_2 \partial p_k} \Delta(c_1 p_k) \right] \\
			& + \frac{\partial^2 H_{\text{eff}}}{\partial p_1 \partial \phi} \Delta(p_2 \phi) - \frac{\partial^2 H_{\text{eff}}}{\partial c_2 \partial \phi} \Delta(c_1 \phi) \\
			& + \frac{\partial^2 H_{\text{eff}}}{\partial p_1 \partial p_\phi} \Delta(p_2 p_\phi) - \frac{\partial^2 H_{\text{eff}}}{\partial c_2 \partial p_\phi} \Delta(c_1 p_\phi).
		\end{split}
		\label{eq:eom_cross_correlation_example}
	\end{equation}
	
	Similar expressions hold for all 28 cross-correlations. The complete system comprises 48 coupled nonlinear ODEs.
	
	\newpage

\bibliographystyle{apsrev4-2}
\bibliography{refs}

%%%%%%%%%%%%%%%%%%%%%%%%%%%%%%%%%%%

\end{document}